\documentclass[suppldata]{interact}

\usepackage{epstopdf}
\usepackage{subcaption}

\usepackage{geometry}
\usepackage{hyperref}
\usepackage[numbers,sort&compress]{natbib}
\bibpunct[, ]{[}{]}{,}{n}{,}{,}
\makeatletter
\def\NAT@def@citea{\def\@citea{\NAT@separator}}
\makeatother

\usepackage{float}
\usepackage{algorithm}
\usepackage{algpseudocode}
\usepackage{afterpage}
\usepackage{capt-of}
\usepackage{enumerate}
\usepackage{multirow}
\usepackage{lscape}
\usepackage{tikz}
\usepackage{color}
\usepackage{circuitikz}
\usepackage{graphicx}

\usepackage{xspace}
\theoremstyle{plain}
\newtheorem{theorem}{Theorem}[section]

\newtheorem{prop}[theorem]{Proposition}

\theoremstyle{definition}
\newtheorem{definition}[theorem]{Definition}

\theoremstyle{remark}
\newtheorem{remark}{Remark}

\begin{document}

	\title{Symmetric Encryption Scheme Based on Quasigroup Using Chained Mode of Operation}
	
	\author{
		\name{Satish Kumar\textsuperscript{a$\ast$}\thanks{*Corresponding author: Satish Kumar. Email: satish.kumar.rs.mat18@itbhu.ac.in}, Harshdeep Singh\textsuperscript{b}\thanks{Harshdeep Singh. Email: harshdeep.sag@gov.in}, Indivar Gupta\textsuperscript{b}\thanks{Indivar Gupta. Email: indivar$\_$gupta@yahoo.com} and Ashok Ji Gupta\textsuperscript{a}\thanks{Ashok Ji Gupta. Email: agupta.apm@itbhu.ac.in}}
		\affil{\textsuperscript{a}Department of Mathematical Sciences, Indian Institute of Technology (BHU), Varanasi, Uttar Pradesh--221005, India; \textsuperscript{b}Scientific Analysis Group, Defence  R$\&$D Organization, Metcalfe House, Delhi--110054, India}
	}
	
	\maketitle
	
	\begin{abstract}
		In this paper, we propose a novel construction for a symmetric encryption scheme, referred as SEBQ which is built on the structure of quasigroup. Utilizing the concepts of chaining like mode of operation, SEBQ is a block cipher with various in-built properties.  This scheme is proven to be resistant against chosen plaintext attack (CPA) and on applying 
		{\it unbalanced Feistel transformation} \cite{desai}, it achieves security against chosen ciphertext attacks (CCA). 
		Subsequently, assessment has been conducted for the randomness of this scheme by running the NIST Statistical test suite be analysing the impact of change in the initial vector, secret key and plaintext on ciphertext through an avalanche effect analysis. 
		Thereafter, a comparative analysis has been done referring the results with existing schemes based on quasigroups \cite{27,38}. 
		Moreover, we also analyse the computational complexity in terms of number of operations needed for encryption and decryption.
		
	\end{abstract}
	
	
	{\bf MSC Classification (2020)}: 20N05, 05B15, 94A60, 68W20. \\

	\maketitle
	\section{Introduction}
	In the late eighteenth century, Euler \cite{3,4} introduced a new theory that delved into the concept of those Latin squares which are pair-wise orthogonal,  termed as MOLS (mutually orthogonal Latin squares). 
	This concept bears a close connection to the structure of quasigroups using combinatorics. 
	Cayley renowned for his exploration of group multiplication tables, demonstrated that such tables could be aptly represented as specially bordered Latin squares \cite{6}. 
	In 1935, Moufang \cite{5} first introduced the term `quasigroup' and later referred to the term `loop' as a quasigroup possessing a unity element. The versatility of quasigroups, with their numerous structures, properties and existence of quasigroups of specific order, enables their application in diverse theories, encompassing  coding theory, cryptography, telecommunications and beyond, for details readers may kindly refer \cite{1,6,8,10,11}. 
	D{\'e}nes and Keedwell, in their work of 1979, 1991 and 2001 \cite{7,8,9}, have conducted an extensive research on  applications of quasigroups in diverse fields like cryptology and coding theory.

	Quasigroup is an algebraic structure that is almost similar to a group, with the difference as the operation in former need not be associative. 
	Hence, we can say every group is a quasigroup not other way around. 
	It is well known that the properties such as closure and inversion of elements in an algebraic structure plays compelling role in the design of various cryptographic primitives. 
	Groups, Rings and Finite fields which are types of associative algebraic structures have been widely employed in the creation of diverse cryptographic primitives, as well as in the development of algebraic codes that can detect and correct errors simultaneously. 
	As already referred, D{\' e}nes and Keedwell \cite{7,8,9} designed various cryptographic primitives using these non-associative structures (quasigroups). 
	
	The security of a cryptographic Hash function NaSHA \cite{nasha}, designed by Markovski et al. in 2008, relied upon factors including the abundance of quasigroup with the same order, the utilization of secret quasigroup operations, large number of isotopes etc.
	The Edon-R encryption scheme, developed by Gligoroski et al. in 2009, \cite{edonr} anchors its security on the intricacies of several factors like  solving the systems of quasigroup operations and Multivariate Quadratic (MQ) equations and finding the order of elements in a given quasigroup. 
	Additionally, several public key cryptographic primitives have been proposed by utilizing the structure of quasigroups.  
	In 2008, Gligoroski et al. \cite{gligoroski2008public} designed a public-key block cipher based on multivariate quadratic quasigroups (MQQs), solving the systems of MQ equations over quasigroup is the backbone of its security.  
	In subsequent years, a digital signature scheme referred as {MQQ-SIG} \cite{gligoroski2011mqq} have been proposed by Gligoroski et al.  and showed experimentally that this scheme is ultra fast and chosen message attack (CMA) resistant. 
	For more details on the construction of public key cryptosystems and other different cryptographic primitives based on quasigroups, readers may kindly refer the survey articles \cite{36} and \cite{1}.    
	
	The exploration of various intriguing areas is unlocked through Latin squares, which serve as the combinatorial structure of quasigroups. These include topics like Latin sub-squares, MOLS and transversals of Latin squares. For an introductory understanding of the theory related to Latin squares and their applications, readers may refer \cite{6,11}. The orthogonal properties of Latin squares and quasigroups  render them suitable for application in coding theory \cite{6,39, 44}. Latin squares also find diverse practical applications, including the design of statistical experiments, construction of error-correcting  telegraph codes, the generation of magic squares, and messages-coding. 
	In 2007, Gligoroski et al. \cite{2} proposed an error correcting code based on  quasigroup transformations. To design such type of codes they utilize an encryption scheme based on quasigroup transformation, producing non-linear and almost random codes. Authors compare their code with Reed-Muller (RM) code of rate $3/16$ that can recover up to 7 errors in 32 bits by showing that the proposed non-linear code can correct even 5 errors in 16 bits which is much better than Reed-Muller code. 
	Belyavskaya, Izbach and Mullen \cite{belyavskaya2005check} proposed a check character system based on using quasigroups and constructed an error-detecting code by utilizing the structure of  T-quasigroups additionally showed that the proposed error detecting code able to detect special types of errors like {\it transposition errors}, {\it twin errors} and {\it phonetic errors}. Subsequently, they apply these error detecting codes on real world applications like ISBN-10.

	\subsection{Quasigroup based encryption schemes}
	Quasigroups find wide applications in the design of various cryptographic primitives including S-boxes \cite{35}, block  ciphers \cite{25,26,27}, stream ciphers \cite{28,29,30}, hash functions \cite{31}, secret sharing schemes \cite{32}, message authentication codes \cite{33} and numerous other areas. For detailed survey on the application of quasigroups in cryptography, readers  are encouraged to  refer the detailed survey articles \cite{1,36}.  
	
	The rise in the use of low memory and computing power devices such as smartphones, various types sensors, palm size computers and  iPads (or tablets)  has created a demand for encryption methods that are well suited to these type of devices. 
	Furthermore, with the growing adoption of cloud services, there is a significant surge in the volume of data being exchanged between these devices, i.e. increasing exponentially \cite{36}. 
	Consequently, there has been  significant attention on developing lightweight cryptographic primitives in recent years. 
	The widespread integration of sensor nodes in low memory devices, RFID tags, microprocessor and low memory chip exemplified pervasive computing, leading to a revolution in modern technology and boost in daily life productivity. 
	However, it also brings security concerns and potential threats to these devices related to the data transmission. 
	To tackle these challenges numerous new cryptographic protocols have been designed and implemented in these types of  resource-constrained scenarios with a particular emphasis on lightweight block ciphers \cite{25,26,27,28,38}. 
	Henceforth, NIST instigated a project aimed at seeking, assessing, and standardizing lightweight cryptographic protocols suitable for these types of  highly constrained (or low memory/ computation) environments. 
	In February 2023, NIST made a public announcement about the formal selection  of the {\it Ascon family} for standardization within the realm of lightweight cryptography. 
	
	In 2012, Battey and Parakh proposed a block encryption scheme based on quasigroup \cite{25}, both with and without cipher block chaining (CBC). They evaluated their scheme's randomness (or entropy) by utilizing NIST statistical test suite and concurrently compared the scheme's performance with AES-256 encryption scheme.
	Their subsequent findings demonstrated the superior performance of their scheme against AES-256.     
	
	In 2021, Tiwari et al. introduced a block cipher {\it INRU: A quasigroup based lightweight block cipher} \cite{27}. 
	They utilized a sequence of string transformations by using the structure of quasigroup in the construction of round functions for both  encryption and decryption algorithms. 
	They  proved that their scheme is robust against various cryptanalytic attacks like linear attack, algebraic attacks and the standard differential attack. Additionally, they performed a comprehensive  statistical analysis using NIST test suite in various mode of operation, including CBC, CFB, OFB and CTR  while comparing their results to AES-128 under the same configuration.
	
	Similarly, Chauhan et al. proposed  a block cipher in 2023, by utilizing the  string transformations based on quasigroup structure referred as BCWST \cite{38}. Likewise, similar to INRU \cite{27}, authors performed a comparison of their design with AES-128 and INRU with the chaining block cipher mode of operation by running the NIST test suite. Additionally,  they proved that their design is secure against various cryptanalytic attacks like standard linear attacks, standard differential attacks, and algebraic attacks.  
	
	This motivates further research in exploring the area of quasigroup based symmetric encryption schemes which are more efficient and have lower complexity with higher security strength.
	
	This paper  focuses on the development of a symmetric encryption scheme based on quasigroup (SEBQ) using a chaining-like mode of operation. Before delving deeper into this topic, we will explore various security aspects related to symmetric encryption schemes. Security analysis entails addressing numerous challenges, such as safeguarding machines against intruders, administering machines to ensure ongoing security, and designing  robust protocols, among others.     
	
	The concept of provable security for public key cryptosystems, proposed by Golwasser and Micali in 1982 \cite{22}, marked a significant advancement. They demonstrated that the security of such systems could be polynomial-time reducible from quadratic residuosity. Subsequently, in 1997, Bellare et al. \cite{12} adapted this concept to the symmetric setting. Prior to this, Shannon introduced the concept of perfect secrecy for cryptosystems, well explained by Stinson \cite{23}. In a perfectly secure cryptosystem, for any two distinct plaintexts $P_1$, $P_2$, and for given legitimate ciphertext $C$ corresponding to any one of plaintexts, the probability of $C$ being the result of encrypting either $P_1$ or $P_2$ is equally likely.
	
	\subsection{Security notions for symmetric encryption scheme}
	For the context of this paper, primarily we discuss two notions of security: chosen-plaintext attack (CPA) and chosen ciphertext attack (CCA). To understand these security of notions, we mainly focus on  conceptual framework discussed by Bellare et al. in \cite{12}, detailed in \cite{13}.
	
	Formally, we model CPA by endowing the adversary $\mathcal{A}$ with the capability to access
	an \textit{encryption oracle}, conceptualized as a ``Black Box" which takes input a message provided by  adversary $\mathcal{A}$ and sends output the encrypted message using key $K$ which is unknown to $\mathcal{A}$.
	In CPA model, adversary $\mathcal{A}$ can access only $\mathcal{E}_K(\cdot)$  and $\mathcal{A}$ can make a query to this oracle by providing input  with a plaintext $m$. In return oracle returns a ciphertext  $c\leftarrow \mathcal{E}_K(m)$ as the output to the adversary $\mathcal{A}$.
	In context of CCA model, the adversary $\mathcal{A}$ possesses access to a decryption oracle $\mathcal{D}_K(c)$ along with the access to an encryption oracle $\mathcal{E}_K(\cdot)$ (Note that adversary $\mathcal{A}$ has no liberty that it can query $\mathcal{D}_K(\cdot)$ on the same ciphertext $c$  output of encryption oracle $\mathcal{E}_K(\cdot)$).  
	The security concepts broadens as  the adversary gains the ability to interact adaptively with the encryption and decryption oracle as needed. 
	
	Consider a symmetric encryption scheme $SE=\{\mathcal{K},\mathcal{E},\mathcal{D}\}$ where $\mathcal{K}$, $\mathcal{E}$ and $\mathcal{D}$ denoted the key generation, encryption  and decryption algorithms respectively. 
	
	\subsubsection{Indistinguishability under chosen-plaintext attack (IND-CPA) and chosen-ciphertext attack (IND-CCA) security} 
	\begin{definition}\cite{13}
		Suppose $\mathcal{A}_{cpa}$ represents an adversary that can access to  encryption oracle $\mathcal{E}_K(\cdot)$ only and $\mathcal{A}_{cca}$ is an adversary that can access both encryption $\mathcal{E}_K(\cdot)$ and decryption  $\mathcal{D}_K(\cdot)$ oracles.
		Now we define the following experiments: for $b\in\{0,1\}$ and $k\in \mathbb{N}$

		\begin{minipage}{0.52\textwidth}
			\begin{algorithm}[H]
				\centering
				\caption{Experiment ${\bf Exp}_{SE,\mathcal{A}_{cpa}}^{ind-cpa-b}(k)$}
				\begin{algorithmic}
					\State 	$K\xleftarrow{R}\mathcal{K}(k)$
					\State	$(x_0,x_1)\leftarrow \mathcal{A}_{cpa}^{\mathcal{E}_K(\cdot)}(k)$
					\State	$c\leftarrow \mathcal{E}_K(x_b)$
					\State 	$b'\leftarrow \mathcal{A}_{cpa}^{\mathcal{E}_K(\cdot)}$
					\State {\bf Return} $b'$
					
				\end{algorithmic}
				\label{algo:ind-cpa}
			\end{algorithm}
		\end{minipage}
		\hfill
		\begin{minipage}{0.52\textwidth}
			\begin{algorithm}[H]
				\centering
				\caption{Experiment ${\bf Exp}_{SE,\mathcal{A}_{cca\textcolor{white}{p}}}^{ind-cca-b}(k)$}
				\begin{algorithmic}
					\State 	$K\xleftarrow{R}\mathcal{K}(k)$
					\State	$(x_0,x_1)\leftarrow \mathcal{A}_{cca}^{\mathcal{E}_K(\cdot),\mathcal{D}_K(\cdot)}(k)$
					\State	$c\leftarrow \mathcal{E}_K(x_b)$
					\State 	$b'\leftarrow \mathcal{A}_{cca}^{\mathcal{E}_K(\cdot),\mathcal{D}_K(\cdot)}$
					\State {\bf Return} $b'$
				\end{algorithmic}
				\label{algo:ind-cca}
			\end{algorithm}
		\end{minipage}
		\\[2mm]
		\noindent Note that $|x_0|=|x_1|$ (i.e. $x_0$ and $x_1$ both have equal lengths) and $\mathcal{A}_{cca}$ cannot query $\mathcal{D}_K(\mathcal{E}_K(\cdot))$
		The advantage of the adversaries $\mathcal{A}_{cpa}$ and $\mathcal{A}_{cpa}$ can be defined as:
		\begin{align*}
			{\bf Adv}_{SE,\mathcal{A}_{cpa}}^{ind-cpa}(k)&=Pr[{\bf Exp}_{SE,\mathcal{A}_{cpa}}^{ind-cpa-1}=1]-Pr[{\bf Exp}_{SE,\mathcal{A}_{cpa}}^{ind-cpa-0}=1]\\		
			{\bf Adv}_{SE,\mathcal{A}_{cca}}^{ind-cca}(k)&=Pr[{\bf Exp}_{SE,\mathcal{A}_{cca}}^{ind-cca-1}=1]-Pr[{\bf Exp}_{SE,\mathcal{A}_{cca}}^{ind-cca-0}=1]
		\end{align*}
		Now the advantage functions for the symmetric encryption scheme $SE$ can be defined as: For any  $t,q_e,\mu_e,q_d,\mu_d\in\mathbb{Z}^{+}$,
		\begin{align*}
			{\bf Adv}_{SE}^{ind-cpa}(k,t,q_e,\mu_e)=\underset{\mathcal{A}_{cpa}}{\max}\{{\bf Adv}_{SE}^{ind-cpa}(k)\}\\
			{\bf Adv}_{SE}^{ind-cca}(k,t,q_e,\mu_e,q_d,\mu_d)=\underset{\mathcal{A}_{cca}}{\max}\{{\bf Adv}_{SE}^{ind-cca}(k)\}
		\end{align*}
		The maximum is taken from all potential adversaries $\mathcal{A}_{cpa}$ and $\mathcal{A}_{cca}$ with time complexity $t$.
		Both adversaries are constrained to make at most $q_e$ queries to the oracle $\mathcal{E}_{K}
		(\cdot)$ and resulting in  $\mu_e$ ciphertexts that both adversaries observe in response to these encryption oracle queries. 
		Additionally, in case of adversary $\mathcal{A}_{cca}$, can make at most $q_d$ queries to the  oracle $\mathcal{D}_{K}(\cdot)$, and as result $\mu_d$ amount of plaintext they see in response to  decryption oracle queries. We say that the symmetric scheme $SE$ is IND-CPA secure (resp. IND-CCA secure), if the advantage function 
		${\bf Adv}_{SE}^{ind-cpa}(\cdot)$ (resp. ${\bf Adv}_{SE}^{ind-cca}(\cdot)$) is negligible  for legitimate adversaries $\mathcal{A}$ with time complexity polynomial in security parameter $k$.  
	\end{definition}
	
	\noindent{\bf Explanation of above experiment}: In the first experiment (\ref{algo:ind-cpa}), a key $K$ is generated by invoking  $\mathcal{K}(k)$, in which input $k$ represents the security parameter. Then,  adversary is provided with the input $k$ and granted access to encryption oracle $\mathcal{E}_K(\cdot)$ (additionally $\mathcal{D}_K(\cdot)$ in case of $cca$ i.e. second experiment (\ref{algo:ind-cca})). 
	A messages tuple ($x_0$, $x_1$) is given to the oracle as input by the adversary, and then oracle selects a random bit $b\in\{0,1\}$ and computes the corresponding ciphertext $c\leftarrow \mathcal{E}_K(x_b)$ and then returns it as output to adversary. 
	The adversary retains ongoing access to  $\mathcal{E}_K(\cdot)$ oracle (additionally, $\mathcal{D}_K(\cdot)$ in case of $cca$), and after performing some calculations, adversary outputs a bit $b'$. 
	The advantage of adversary can be computed as the difference between the probabilities of correctly and incorrectly guessing the bit $b'$ linked to the encryption oracles input $b$.
	
	\subsubsection{Left-or-Right indistinguishability under chosen-plaintext attack (LOR-CPA) and chosen-ciphertext attack (LOR-CCA)}
	
	\begin{definition}\cite{12}
		Suppose $\mathcal{A}_{cpa}$ represents  an adversary that can access only encryption oracle $\mathcal{E}_{K}(LR(\cdot,\cdot,b))$ and $\mathcal{A}_{cca}$ is an adversary  that can access  both the encryption $\mathcal{E}_{K}(LR(\cdot,\cdot,b))$ and decryption  $\mathcal{D}_K(\cdot)$ oracles. Now we define the following experiments:  for $b\in\{0,1\}$ and $k\in \mathbb{N}$

		\begin{minipage}{0.52\textwidth}
			\begin{algorithm}[H]
				\centering
				\caption{Experiment ${\bf Exp}_{SE,\mathcal{A}_{cpa}}^{lor-cpa-b}(k)$}
				\begin{algorithmic}
					\State 	$K\xleftarrow{R}\mathcal{K}(k)$
					\State 	$d\leftarrow\mathcal{A}_{cpa}^{\mathcal{E}_{K}(LR(\cdot,\cdot,b))}(k)$
					\State {\bf Return} $d$.
				\end{algorithmic}
			\end{algorithm}
		\end{minipage}
		\hfill
		\begin{minipage}{0.52\textwidth}
			\begin{algorithm}[H]
				\centering
				\caption{Experiment ${\bf Exp}_{SE,\mathcal{A}_{cca\textcolor{white}{p}}}^{lor-cca-b\textcolor{white}{p}}(k)$}
				\begin{algorithmic}
					\State 	$K\xleftarrow{R}\mathcal{K}(k)$
					\State 	$d\leftarrow\mathcal{A}_{cca\textcolor{white}{p}}^{\mathcal{E}_{K}(LR(\cdot,\cdot,b)),\mathcal{D}_{K}(\cdot)}(k)$
					\State {\bf Return} $d$.
				\end{algorithmic}
			\end{algorithm}
		\end{minipage}
		\\[2mm]
		\noindent Note that $|x_0|=|x_1|$ and  $\mathcal{A}_{cca}$ cannot query $\mathcal{D}_K(\mathcal{E}_K(LR(\cdot,\cdot,b)))$.  The advantage of the adversaries $\mathcal{A}_{cpa}$ and $\mathcal{A}_{cpa}$ is defined as:
		\begin{align*}
			{\bf Adv}_{SE,\mathcal{A}_{cpa}}^{lor-cpa}(k)=Pr[{\bf Exp}_{SE,\mathcal{A}_{cpa}}^{lor-cpa-1}=1]-Pr[{\bf Exp}_{SE,\mathcal{A}_{cpa}}^{lor-cpa-0}=1]\\
			{\bf Adv}_{SE,\mathcal{A}_{cca}}^{lor-cca}(k)=Pr[{\bf Exp}_{SE,\mathcal{A}_{cca}}^{lor-cca-1}=1]-Pr[{\bf Exp}_{SE,\mathcal{A}_{cca}}^{lor-cca-0}=1]
		\end{align*}
		Now the advantage functions for the symmetric encryption scheme $SE$ can be defined as: For any $t,q_e,\mu_e,q_d,\mu_d\in\mathbb{Z}^{+}$,
		\begin{align*}
			{\bf Adv}_{SE}^{lor-cpa}(k,t,q_e,\mu_e)=\underset{\mathcal{A}_{cpa}}{\max}\{{\bf Adv}_{SE}^{lor-cpa}(k)\}\\
			{\bf Adv}_{SE}^{lor-cca}(k,t,q_e,\mu_e,q_d,\mu_d)=\underset{\mathcal{A}_{cca}}{\max}\{{\bf Adv}_{SE}^{lor-cca}(k)\}
		\end{align*}
		The maximum is taken from all potential adversaries $\mathcal{A}_{cpa}$ and $\mathcal{A}_{cca}$ with time complexity $t$.
		Both adversaries are constrained to make at most $q_e$ queries to the oracle $\mathcal{E}_{K}(LR(\cdot,\cdot,b))$ and resulting in  $\mu_e$ ciphertexts that both adversaries observe in response to these encryption oracle queries. Additionally, $\mathcal{A}_{cca}$ can make at most $q_d$ queries to the  oracle $\mathcal{D}_{K}(\cdot)$, and as result $\mu_d$ amount of plaintext they sees in response to  decryption oracle queries.  We say  that the scheme $SE$ is  LOR-CPA secure (resp. LOR-CCA secure) if the advantage function ${\bf Adv}_{SE}^{lor-cpa}(\cdot)$ (resp. ${\bf Adv}_{SE}^{lor-cca}(\cdot)$) is negligible for legitimate adversaries $\mathcal{A}$ time complexity polynomial in $k$.

	\end{definition}

	\subsubsection{Related works}
	Several practical  modes of operations have been proposed after realizing the practicality of block ciphers and these modes are divided into mainly three different categories: confidentiality modes, authenticated modes and authenticated-encryption modes.  Some examples of confidentiality and authenticated mode of  operations are:  Electronic code book (ECB),  Cipher block chaining (CBC),   Cipher feedback (CFB), Output feedback (OFB), Counter (CTR),  Cipher block chaining-message authentication code (CBC-MACs), Hash message authentication code (HMAC) and Galois message authentication code (GMAC). The authenticated-encryption mode of operations ensures both the confidentiality and authenticity of the ciphertext. Examples of the authenticated encryption scheme include: Galois counter mode (GCM) \cite{14} and Counter with Cipher Block Chaining-Message (CCM) Authentication Code \cite{14}. For detailed survey on block cipher using different modes of operation, readers may refer \cite{14}.  
	
	In past, various confidentiality modes have been proposed like ECB, CBC, CFB etc. 
	We discuss some of the chaining like confidential modes with the bounds on advantage function ${\bf Adv}$ using advantage of pseudorandom functions. 
	\begin{itemize}
		\item {\bf Electronic code book (ECB)}: In FIPS 1981 \cite{15},   various modes of operation for Data Encryption Standards (DES) are described and ECB is one of them. It  is a deterministic  mode of operation, in which the blocks of messages are encrypted independently. This vulnerability makes ECB susceptible to chosen plaintext attack (CPA).
		
		The lack of diffusion in ciphertext is the main drawback for this mode. ECB encryption scheme transforms identical plaintext blocks into the identical ciphertext blocks and fails to effectively conceal the data patterns. For instance, it is easy to differentiate  the encryption of a plaintext containing two identical blocks from the encryption of a plaintext that consists of two distinct blocks.

		\item {\bf Cipher block chaining (CBC) }: In 1976, Ehrsam  et al. introduced the CBC mode of operation \cite{16}. 
		In this mode, unlike ECB, an encrypted form of initialization vector is XORed with the first block of plaintext, and then each consecutive plaintext block is XORed with the preceding ciphertext block before the encryption process. The initial vector guarantees the uniqueness of ciphertext when same message is encrypted more than once. In \cite{12}, Bellare et al. did a comprehensive  security analysis for the CBC scheme and proved that it is LOR-CPA secure.
		
		The primary drawbacks includes that the encryption scheme is sequential, and before encryption user must ensure that 
		the message is padded to a scaler multiple of the cipher block size. Additionally, in the CBC mode, even a single bit change in the plaintext or initial vector can impact all the subsequent ciphertext blocks.
		
		In \cite{12}, Bellare et al. gave lower and upper bound on advantage $({\bf Adv})$ of CBC scheme using the \textit{random permutations}, i.e. for $R=\{0,1\}^l$, $\mu_e\le l\cdot 2^{1\slash 2}$ and $q_e=\mu_e\slash l$
		\begin{equation*}
			0.316\left(\frac{\mu_e^2}{l^2}-\frac{\mu_e}{l}\right)\cdot \frac{l}{2^l}\le {\bf Adv}_{CBC[R]}^{lor-cpa}(\cdot,t,q_e,\mu_e)\le \left(\frac{\mu_e^2}{l^2}-\frac{\mu_e}{l}\right)\cdot \frac{l}{2^l}.
		\end{equation*}
		In \cite{12}, Bellare et al. also gave an upper bound on the advantage ${\bf Adv}$ of CBC scheme compared to the \textit{pseudorandom  permutations}. Suppose $F$ belongs to pseudorandom  permutations family with length $l$. For any $t,q_e$ and $\mu_e=ql$,
		\begin{equation*}
			{\bf Adv}_{CBC[F]}^{lor-cpa}(\cdot,t,q_e,\mu_e)\le 2\cdot {\bf Adv}_F^{prp}(t,q)+\frac{q^2}{2^{l+1}}+\left(\frac{\mu_e^2}{l^2}-\frac{\mu_e}{l}\right)\cdot \frac{1}{2^l}.
		\end{equation*}
		\begin{prop}[\cite{20}]
			Consider a $CBC$ scheme $SE=\{\mathcal{K},\mathcal{E},\mathcal{D}\}$ with random initial vector $IV$ and $\mathcal{E}:\mathcal{K}\times \{0,1\}^n\rightarrow \{0,1\}^n$ is an encryption map.
			Then
			\begin{equation*}
				{\bf Adv}_{SE}^{ind-cca}(t,1,n,1,2n)=1
			\end{equation*} 
			for $t=\mathcal{O}(n)$, in addition to the time required for a single application of $F$, the advantage of this adversary is 1 despite utilizing minimal resources: merely one query to each oracle. This unequivocally indicates that the scheme is IND-CCA insecure.  
		\end{prop}
		
		\item {\bf Cipher feedback block (CFB)}: In 2002, Alkassar et al. introduced the CFB mode of operation \cite{17}. In CFB scheme, the first block of plaintext is XORed with the encrypted form of initial vector $IV$, while the succeeding blocks of plaintext are XORed with the encrypted form of the previous block cipher.
		NIST SP800-38A establishes the definition of CFB with a specified bit-width \cite{18}. CFB mode requires a parameter, denoted by $s$, where $1\le s\le b$ where $b$ is block length. In  the description  of  CFB mode, each plaintext and ciphertext segment is $s$ bit long. Sometimes, the value of $s$ is also incorporated into the mode's nomenclature, such as CFB mode, CFB-8 mode, CFB-64 mode or CFB-128 mode of operation.
		
		The disadvantages of CFB are identical to CBC scheme.  In \cite{19}, Wooding gave the bound on advantage of CFB scheme using the pseudorandom function $F$. For arbitrary $t_e,q_e$  and $\mu_e$,
		\begin{equation*}
			{\bf Adv}^{lor-cpa}(CFB,t_e,q_e,\mu_e)\le 2\cdot {\bf Adv}^{prf}(F,t_e+qt_f,q)+\frac{q(q-1)}{2^l}
		\end{equation*} 
		where $q=\lfloor (\mu_e+q_e(t-1))\slash t\rfloor+n_e$, $t_f$ is some small constant, $l$ is the length of message block and $prf$ represent pseudorandom function family.
		
		\item {\bf Output Feedback block (OFB)}:  The OFB mode of operation is seen as the another variant of the CBC. In OFB mode \cite{14}, first block of the plaintext is XORed with the encrypted form of initial vector  and for the second block cipher, encrypted form of initial vector is passed to the second encryption block and then the second plaintext block XORed with the output of the encryption block.   
		
		OFB mode achieves CPA security when the encryption function $F$ is a pseudorandom function. This mode, while requiring sequential encryption, offers the advantage over  CBC mode  that most of the computation, specifically the computation  of the  pseudorandom stream, can be performed independently, irrespective of the  specific  message intended be encryption.   
		
		In \cite{19}, Wooding gave the bound on advantage of OFB scheme using the pseudorandom function $F$. For any $t_e,q_e$  and $\mu_e$,
		\begin{equation*}
			{\bf Adv}^{lor-cpa}(OFB,t_e,q_e,\mu_e)\le 2\cdot {\bf Adv}^{prf}(F,t_e+qt_f,q)+\frac{q(q-1)}{2^l}
		\end{equation*} 
		where $q=\lfloor (\mu_e+q_e(t-1))\slash t\rfloor+n_e$, $t_f$ is some small constant, $l$ is the length of message block and $prf$ represent pseudorandom function family. 
		
		\item {\bf Counter (CTR)}: In 1979, Diffie and Hellman  introduced the CTR mode of operation. This mode is straightforward counter-based implementation of a block cipher. With each encryption operation, a counter-initiated value is XORed with the plaintext and then perform the encryption process to produce the ciphertext blocks. The use of a counter value for each block helps avoid establishing a relationship between  plaintext and ciphertext.
		
		The main disadvantage of CTR mode is its dependence on a synchronous counter during both encryption and decryption process. The recovery of plaintext is  erroneous when synchronization is lost. 
		
		In \cite{12}, Bellare et al. gave the bound on advantage  of CTR mode using the pseudorandom function $F$ with input length $l$ and output length $L$. For any $t,q_e$ and $\mu_e=\min(q'L,L2^l)$,
		\begin{equation*}
			{\bf Adv}_{CTR[F]}^{lor-cpa}(\cdot,t,q_e,\mu_e)\le 2\cdot {\bf Adv}_F^{prf}(t,q')
		\end{equation*}  
		
		\begin{prop}\cite{20}
			Consider a counter based encryption scheme  $SE=\{\mathcal{K},\mathcal{E},\mathcal{D}\}$ with random counter $CTR$. Let $F:\mathcal{K}\times \{0,1\}^n\rightarrow \{0,1\}^n$ be the corresponding family of functions. Then the advantage function of the $SE$ is:
			\begin{equation*}
				{\bf Adv}_{SE}^{ind-cca}(t,1,l,1,n+l)=1.
			\end{equation*} 
			As we can observe that the advantage of the adversary is 1 using just one query to each encryption and decryption oracle with the time complexity as $t=\mathcal{O}(n+l)$ plus one operation of $F$. This clearly represents that the counter based encryption scheme $SE$ with random $CTR$ is  IND-CCA insecure.
		\end{prop}
	\end{itemize}
	
	\subsection{\bf Main contribution of the paper}
	\begin{itemize}
		\item We propose a novel construction of symmetric encryption scheme SEBQ based on the structure of quasigroup. 
		SEBQ employs an in-built chaining-like mode of operation. In this scheme, transformed initial vector is utilized to encrypt the subsequent message block rather than using  ciphertext of the preceding block.    
		
		\item For security assessment of SEBQ, we prove that this scheme is IND-CPA secure and after applying {unbalanced Feistel transformation}, we establish its security strength to IND-CCA2 secure.
		
		\item Our analysis also involve assessing the performance of SEBQ scheme in contrast to the existing encryption schemes like INRU \cite{27}, BCWST \cite{38}, AES-128. This assessment primarily focuses on evaluating the  randomness of the generated sequences by running the NIST statistical test suite (NIST-STS). Additionally, we  investigate the avalanche effect concerning  the secret key, plaintext and random initial vector within the SEBQ scheme.
		
		\item Finally we perform computational analysis of SEBQ focusing on computational complexity in terms of number of operations required. Subsequently, we establish a relation with the order of Latin square to achieve 128-bit and 256-bit security against known-ciphertext attack. 
	\end{itemize}

	\noindent{\bf Organization of the paper}\\
	This article is structured into several sections:
	First, in Section \ref{sec:prelims} we give the preliminaries of quasigroups that are required to understand this paper and the string transformations based on structure of quasigroups. 
	Section \ref{sec:mobq} is dedicated to the construction of SEBQ using chained like mode of operations in which instead of ciphertext transformed random vector will be passed to next block for the encryption of succeeding blocks of plaintext.
	Section \ref{sec:security analysis of MOBQ} is committed for the security analysis of SEBQ scheme, in which we prove main results of this paper and we also do the randomness testing of the SEBQ scheme. In Section \ref{sec:analysisof scheme}, we find out the computational complexity of the SEBQ scheme. Finally, In Section \ref{sec:conclusion}, we draw the conclusions of the paper.

	\section{Preliminaries}\label{sec:prelims}
	This section contains a concise overview related to the theory of quasigroups, the definition of symmetric encryption scheme and the string transformations based on quasigroup. For detailed theory of quasigroup and the symmetric encryption schemes, readers may refer \cite{1,12,20,23}.

	Consider a finite set $Q$  and a binary operation $*:Q^2\rightarrow Q$ possessing the following property: for all $a,b\in Q$ there exist unique $x,y\in Q$ satisfying the equations $a*x=b$ and $y*a=b$. Then, $(Q,*)$ is known as quasigroup.
	
	For a given quasigroup $(Q,*)$, the parastrophe or adjoint binary operation `$\backslash$' can be derived from  the given binary operation  `$*$' by utilizing the following equation: 
	\begin{equation}\label{eq:parastropheofQ}
		x*y=z\Leftrightarrow y=x\backslash z	
	\end{equation} 
	The algebra $(Q,*,\backslash)$ satisfies the following identities:
	$$
	x\backslash(x*y)=y~ \text{and}~ x*(x\backslash y)=y,
	$$ and $(Q,\backslash)$ is also a quasigroup.
	
	\begin{definition}\cite{6}\label{def:latin square}
		A  Latin square is an array of size $n\times n$ filled with $n$ distinct symbols, ensuring that  each symbol appears precisely once in each row and once in each column. 
	\end{definition}

	Every quasigroup $(Q=\{x_1,x_2,\dots,x_n\},*)$ can be represented by a Latin square $L$ of size $n^2$. For example, for the set $Q=\{1,2,3,4,5\}$, a Latin square and its corresponding parastrophe's Latin square are represented in Table \ref{table:quasigroup}:

	\begin{table}[H]
		\centering
		
		\begin{subtable}[H]{0.48\textwidth}
			\centering
			\begin{tabular}{l|lllll}
				
				$*$ & $1$ & $2$ & $3$ & $4$ & $5$\\
				\hline
				$1$ & $1$ & $2$ & $3$ & $4$ & $5$\\
				
				$2$ & $2$ & $1$ & $4$ & $5$ & $3$\\
				
				$3$ & $3$ & $5$ & $1$ & $2$ & $4$\\
				
				$4$ & $4$ & $3$ & $5$ & $1$ & $2$\\
				
				$5$ & $5$ & $4$ & $2$ & $3$ & $1$
				
			\end{tabular}
			
		\end{subtable}
		\hfill
		\begin{subtable}[H]{0.48\textwidth}
			\centering
			\begin{tabular}{l | l llll}
				$\backslash$ & $1$ & $2$ & $3$ & $4$ & $5$\\
				\hline
				$1$ & $1$ & $2$ & $3$ & $4$ & $5$\\
				
				$2$ & $2$ & $1$ & $5$ & $3$ & $4$\\
				
				$3$ & $3$ & $4$ & $1$ & $5$ & $2$\\
				
				$4$ & $4$ & $5$ & $2$ & $1$ & $3$\\
				
				$5$ & $5$ & $3$ & $4$ & $2$ & $1$

			\end{tabular}
			
		\end{subtable}
		
		\caption{Latin squares  corresponding to quasigroup $(Q,*)$ and its parastrophes $(Q,\backslash)$  }
		
		\label{table:quasigroup}
	\end{table}

	\begin{definition}\cite{40}\label{def:permanentofmatrix}
		Consider a matrix $A=(a_{i,j})$ of order $n\times n$.
		Then, the  permanent of $A$ is defined  as:
		\begin{equation*}
			perm(A)=\sum_{\tau\in S_n}^{}\prod_{j=1}^{n}a_{j,\tau(j)}.
		\end{equation*}
		The summation is running for all $\tau$ belongs to symmetric group $S_n$
	\end{definition}
	
	The  enumeration of Latin square has a rich history. Firstly, in late eighteenth century Euler introduced this problem in his work \cite{3}. Research have shown that for order 2 and 3 there exist one Latin square, for order 4 and 5 there exist four and fifty six Latin squares respectively \cite{4}. Subsequently,  in 1990, Galina et al. \cite{45} gave  the count of Latin squares of order 8; while in 2011 A. Hulpke et al. \cite{47} gave the count of Latin squares of order 11. Recently, in 2019
	Eduard Vatutin et al. gave a general formula to find out the enumeration of Latin square of order $n$
	\cite{46}. Furthermore, by volunteer counting one may compute the number of  isotopy classes of diagonal Latin squares with small order. 
	
	The cardinality of Latin square plays a significant role in theoretical security of quasigroup-based cryptographic protocols. However, calculating the  number of Latin square \cite{shao1992formula} poses a challenge because there is exponential growth in the number of terms involved. Consequently, extensive research efforts have been dedicated to estimating the  number of Latin square for specific order by employing methods like volunteer counting \cite{3,4,45,46,47} or utilized specialized  software like Singular/ SageMath \cite{46,43}.

	In \cite{shao1992formula}, Shao and Wei determined a formula to calculate the  number of Latin squares. Suppose $\tau_0(A)$ represents the total number of zeroes in matrix $A$ and $perm(A)$ is the permanent of $A$ defined as (\ref{def:permanentofmatrix}), the respective formula is described below:
	\begin{theorem}
		Let $B_n=\{(b_{i,j})_{n\times n}\mid b_{i,j}\in\{0,1\}\}$ be  the set of all $n\times n$ sized matrices. Then number of Latin squares is:
		\begin{equation}\label{eq:latinsquare}
			L(n)=n!\sum_{A\in B_n}^{}(-1)^{\tau_0(A)}\binom{perm(A)}{n}
		\end{equation}
	\end{theorem}
	
	\begin{remark}
		Let $B_{k\times n}=\{(b_{i,j})_{k\times n}\mid b_{i,j}\in\{0,1\}\}$ be the set of $k\times n$ matrices. Then the formula to calculate the number of Latin rectangles is:
		\begin{equation}\label{eq:lainsquare1}
			L({k\times n})=n!\sum_{A\in B_{k\times n}}(-1)^{\tau_0(A)}\binom{perm(A)}{n}~~~~\text{for}~ k\le n
		\end{equation}
	\end{remark}
	
	Above formulae (\ref{eq:latinsquare}) and (\ref{eq:lainsquare1}) are not easy for computation, however D{\'e}nes and Keedwell \cite{6}  used them to compute bounds on number of Latin squares, i.e.
	\begin{equation}\label{eq:boundsonlatinsquare}
		\prod_{j=1}^{n}(j!)^{n/j}\ge L(n)\ge \frac{(n!)^{2n}}{n^{n^2}} .
	\end{equation}
	For specific values of $n=2^l $, where $l=7,8$, the approximate number of $L(n)$ is estimated as:
	\begin{align*}
		0.164\times 10^{21091}\ge L(128)\ge 0.337\times 10^{20666}\\
		0.753\times 10^{102805}\ge L(256)\ge 0.304\times 10^{101724}
	\end{align*}
	We give a table contains exact values of Latin square $L(n)$ with $n$-number of symbols.   
	\begin{table}[H]
		\centering
		\caption{Number of Latin squares of size $n$.}
		\begin{tabular}{|c|c|}
			\hline
			$n$ & Number of Latin squares of size $n$\\
			\hline 
			$1$ & $1$\\
			\hline
			$2$ & $2$\\
			\hline
			$3$ & $12$\\
			\hline
			$4$ & $576$\\
			\hline 
			$5$ & $161,280$\\
			\hline
			$6$ & $812,851,200$\\
			\hline
			$7$ & $61,479,419,904,000$\\
			\hline
			$8$ & $108,776,032,459,082,956,800$\\
			\hline
			$9$ & $5,524,751,496,156,892,842,531,225,600$\\
			\hline
			$10$ & $9,982,437,658,213,039,871,725,064,756,920,320,000$\\
			\hline
		\end{tabular}
		
		\label{table:latinsquare}
	\end{table}
	
	In order to set-up an encryption scheme, we make use of the quasigroup $(Q,*)$ based transformations on $Q^l\times Q^n$. This is formally defined below: 
	
	\begin{definition}
		\cite{2}\label{def:e-transformation}{\bf ($\mathfrak{e}$-transformation)}
		Consider a quasigroup $(Q,*)=(\mathbb{F}_2^k,*)$.  
		For a given $b\in Q$, we define $F_b:Q\rightarrow Q$ as $F_b(a)=b*a$. 
		Also, we define $f:Q^n\rightarrow Q^n$ as $f(b_1b_2\dots b_n)=d_1d_2\dots d_n$, where each $d_i=b_i$ for $i=\overline{1,n-1}$ and $d_n=b_1 + b_2+\cdots + b_n$  (where $\overline{1,n-1}$ represents integers ranging from $1$ to $n-1$).
		Using these functions, we define the following:
		\begin{enumerate}[$(i)$]\label{eq:e-stringtrans}
			\item For a given $\beta=b_1b_2\dots b_n\in Q^n$, we define a map $F_{\beta}:Q\rightarrow Q$ as $F_{\beta}=F_{b_n}\circ F_{b_{n-1}}\circ\dots \circ F_{b_1}$.
			
			\item For a given $a\in Q$, we define a map $f_a^n:Q^n\rightarrow Q^n$ as $f_a^n(b_1b_2\dots b_n)=d_1d_2\dots d_n$, where each component maps differently as $d_i=F_{b_i}\circ F_{b_{i-1}}\circ \cdots \circ F_{b_1}(a)$, $i=\overline{1,n}$; additionally, we define $F_a^n:Q^n\rightarrow Q^n$ as $F_a^n=f\circ f_a^n$.
			
			\item Finally, we define $\mathfrak{e}$-transformation $F^{l,n}:Q^l\times Q^n\rightarrow Q^l\times Q^n$ as $F^{l,n}(a_1a_2\dots a_l,~b_1b_2\dots b_n)=(c_1c_2\dots c_l,~ d_1d_2\dots d_n)$, where each component is computed as:
			$$
			c_i=F_{\delta_{i-1}}(a_i),~\delta_i=F_{a_i}^n(\delta_{i-1}) \text{ for $i=\overline{1,l}$}
			$$   assuming $\delta_0=b_1b_2\dots b_n$, and $d_1d_2\dots d_n = \delta_l$.
			
		\end{enumerate}

	\end{definition}  
	\begin{definition}\cite{2}\label{def:d-transformation}
		{\bf ($\mathfrak{d}$-transformation)}
		Consider a quasigroup $(Q,*)=(\mathbb{F}_2^k,*)$ and its corresponding parastrophe $\backslash$. 
		For a given $b\in Q$, we define $G_b:Q\rightarrow Q$ as $G_b(c)=b\backslash c$. Also, we define $f:Q^n\rightarrow Q^n$ as $f(b_1b_2\dots b_n)=d_1d_2\dots d_m$ where each $d_i=b_i$  for $i=\overline{1,n-1}$ and $d_n=b_1+b_2+\dots + b_n$. Using these functions, we define the following:
		\begin{enumerate}[$(i)$]\label{eq:d-stringtrans}
			
			\item For a given $\beta=b_1b_2\dots b_n\in Q^n$, we define a map  $G_{\beta}:Q\rightarrow Q$  as $G_{\beta}=G_{b_1}\circ G_{b_2}\circ \dots \circ G_{b_n}$.
			
			\item For a given $c\in Q$, we define a map $g_c^n:Q^n\rightarrow Q^n$ as $g_c^{n}(b_1b_2\dots b_n )=d_1d_2\dots d_m$  and each component map differently as $d_{n-i+1}=G_{b_n}\circ G_{b_{n-1}}\circ \dots \circ G_{b_{n-i+1}}(c)$, $i=\overline{1,n}$; additionally, we define  $G_c^n:Q^n\rightarrow Q^n$  as $G_c^n=f\circ g_c^n$.
			
			\item Finally, we define $\mathfrak{d}$-transformation $G^{l,n}:Q^l\times Q^n\rightarrow Q^l\times Q^n$ as \\$G^{l,n}(c_1c_2\dots c_l,b_1b_2\dots b_n)=(a_1a_2\dots a_l,d_1d_2\dots d_n)$ where each component computed  as:
			$$
			a_i=G_{\delta_{i-1}}(c_i),~\delta_i=G_{c_i}^n(\delta_{i-1})
			$$   assuming  $\delta_0=b_1b_2\dots b_n$ and $d_1d_2\dots d_n=\delta_l$.
		\end{enumerate}
		
	\end{definition}
	
	\begin{theorem}\label{thm:useofstringtrans}\cite{2} From the functions defined in Definition \ref{def:e-transformation} and Definition \ref{def:d-transformation}, we have the following:
		\begin{enumerate}[$(i)$]
			\item $G_{\beta}^l(F_{\beta}^l(\alpha))=\alpha~\text{and}~ F_{\beta}^l(G_{\beta}^l(\gamma))=\gamma$,
			
			\item $F_{\beta}^l$ and $G_{\beta}^l$ are permutations of $Q^l$.
		\end{enumerate}
	\end{theorem}

	{\bf Symmetric encryption scheme}: In this paper, our objective is to  construct a symmetric encryption scheme $SE=\{\mathcal{K},\mathcal{E},\mathcal{D}\}$ that comprises of three algorithms: Key Generation Algorithm, Encryption Algorithm and Decryption Algorithm. 
	\begin{enumerate}[$(i)$]
		\item {\it Key Generation}: In Key Generation Algorithm $\mathcal{K}$, it  accepts a security parameter $k\in \mathbb{N}$ as input and returns a key secret key $K$ as output, denoted as $K\xleftarrow{R}\mathcal{K}(k)$. 
		
		\item {\it Encryption algorithm}: The Encryption Algorithm $\mathcal{E}_K:Q\rightarrow Q$ can be either randomized or stateful. When  instantiated with the key $K$, it takes a plaintext $M\in Q^l$ as input and returns a ciphertext $C$, denoted as $C\xleftarrow{R}\mathcal{E}_K(M)$.
		
		\item {\it Decryption algorithm}: The decryption algorithm $\mathcal{D}_K$ is both deterministic and stateless. It initializes with a key $K$ and processes a string $C$ to return the corresponding plaintext $x\in Q^l$, denoted as $x\leftarrow\mathcal{D}_K(C)$. It is important to ensure that the design of $\mathcal{E}$ and $\mathcal{D}$ guarantees that $\mathcal{D}_K(\mathcal{E}_K(x))=x$ holds for all $x\in Q^l$.
		
	\end{enumerate}

	\section{Symmetric encryption scheme based on quasigroup (SEBQ)}\label{sec:mobq}
	In this section, we design a symmetric encryption scheme based on the $\mathfrak{e}$ and $\mathfrak{d}$-transformations using Definition \ref{def:e-transformation} and \ref{def:d-transformation} of the quasigroup by utilizing chaining-like mode of operation. We design a block cipher of size $k$-bit long blocks and the operation of quasigroup $*$ as  secret of our encryption scheme.  
	
	Consider a set $Q=\mathbb{F}_2^k$ for some positive integer $k$. Now, $a\in Q$ implies $a$ can be written as $k$-tuple of elements from set $\{0,1\}$. 
	In the SEBQ scheme, three algorithms have been put forth, including {\it Key Generation algorithm}, an {\it Encryption algorithm} and the {\it Decryption algorithm}. The security of  SEBQ  depends on the number of quasigroups of particular order.  
	
	\subsection{Key Generation}
	
	Generate a random Latin square of size $2^k\times 2^k$ by giving an input a parameter $k\in\mathbb{N}$, utilizing an algorithm proposed by Artamonov et al. \cite{43}. The generated Latin square is being treated as a secret key for SEBQ scheme and the  dimension of the key space is $|L(n)|$ for $n=2^k$ where as $|L(n)|$ can be calculated by utilizing (\ref{eq:latinsquare}). To compute the parastrophe of the corresponding Latin square, we utilize (\ref{eq:parastropheofQ}), which states that the knowledge of $(Q,*)$ or combinatorial equivalent Latin square is equivalent to the knowledge of $(Q,\backslash)$.

	\subsection{Encryption Process}\label{enc_sub}
	Consider $M=(m_1,m_2,\dots,m_l)\in (\mathbb{F}_2^k)^l$, where each $m_i\in \mathbb{F}_2^k$, i.e. each block $m_i$ is $k$-bit long and $M$ is $kl$-bit long plaintext. 
	In order to begin the encryption process, user needs to consider an initial vector $R^{(1)}=r_1^{(1)},r_2^{(1)},\dots, r_n^{(1)}$, i.e. a $kn$-bit long string, where each $r_i^{(1)}\in Q$.
	
\begin{figure}[H]
		\resizebox{\columnwidth}{!}{
			\begin{tikzpicture}
				\node at (1,8) {${r_1}^{(J)}$};
				\node at (2.5,8) {${r_2}^{(J)}$};
				\node at (4,8) {${r_3}^{(J)}$};
				\node at (11,8) {${r_{n-1}}^{(J)}$};
				\node at (13,8) {${r_n}^{(J)}$};
				\node[scale=1.5] at (1,6.7) {$*$};
				\node[scale=1.5] at (2.5,6.7) {$*$};
				\node[scale=1.5] at (4,6.7) {${*}$};
				\node[scale=1.5] at (11,6.7) {$*$};
				\node[scale=1.5] at (13,6.7) {$*$};
				\node at (1,5.1) {${r_1}^{(J+1)}$};
				\node at (2.5,5.1) {${r_2}^{(J+1)}$};
				\node at (4,5.1) {${r_3}^{(J+1)}$};
				\node at (11,5.1) {${r_{n-1}}^{(J+1)}$};
				\node at (13,5.1) {${r_n}^{(J+1)}$};
				\node at (1,3.4) {${r_1}^{(J+1)}$};
				\node at (2.5,3.4) {${r_2}^{(J+1)}$};
				\node at (4,3.4) {${r_3}^{(J+1)}$};
				\node at (11,3.4) {${r_{n-1}}^{(J+1)}$};
				\node at (13.7,3.4) {$\bigoplus\limits_{i=1}^{n}{r_{i}}^{(J+1)}\xrightarrow{}{r_{n}}^{(J+1)}$};
				\draw (0.2,7.5)--(13.8,7.5)--(13.8,8.5)--(0.2,8.5)--(0.2,7.5);
				\draw (0.2,4.7)--(13.8,4.7)--(13.8,5.7)--(0.2,5.7)--(0.2,4.7);
				\draw (0.2,3)--(15.5,3)--(15.5,4)--(0.2,4)--(0.2,3);
				\draw[thick] (1,7.7)--(1,7);
				\draw[thick] (2.5,7.7)--(2.5,7);
				\draw[thick] (4,7.7)--(4,7);
				\draw[thick] (11,7.7)--(11,7);
				\draw[thick] (13,7.7)--(13,7);
				\draw[thick] [-latex] (1,4.7)--(1,3.8);
				\draw[thick] [-latex] (2.5,4.7)--(2.5,3.8);
				\draw[thick] [-latex] (4,4.7)--(4,3.8);
				\draw[thick] [-latex] (11,4.7)--(11,3.8);
				\draw[thick] [-latex] (13,4.7)--(13,3.8);
				\draw[thick] [-latex] (1.2,5.4)--(2.4,6.6);
				\draw[thick] [-latex] (2.7,5.4)--(3.9,6.6);
				\draw[thick] [-latex] (11.2,5.4)--(12.9,6.6);
				\draw[thick] [-latex] (1,6.5)--(1,5.5);
				\draw[thick] [-latex] (2.5,6.5)--(2.5,5.5);
				\draw[thick] [-latex] (4,6.5)--(4,5.5);
				\draw[thick] [-latex] (11,6.5)--(11,5.5);
				\draw[thick] [-latex] (13,6.5)--(13,5.5);
				\node[scale=1] at (-0.5,8) {$R^{(J)}$};
				\node[scale=1] at (-0.5,3.4) {$R^{(J+1)}$};
				\node[scale=1] at (0,6.7) {$m_J$};
				\draw[thick] (0.3,6.7)--(0.8,6.7);
				\draw[thick,loosely dashed] (4.4,8)--(10.3,8);
				\draw[thick,loosely dashed] (4.6,5.1)--(10.1,5.1);
				\draw[thick,loosely dashed] (4.6,3.4)--(10.1,3.4);
				
				
		\end{tikzpicture}	}
		\caption{Transformation of initial vector for encryption function}
		\label{fig:transformvector}
	\end{figure}
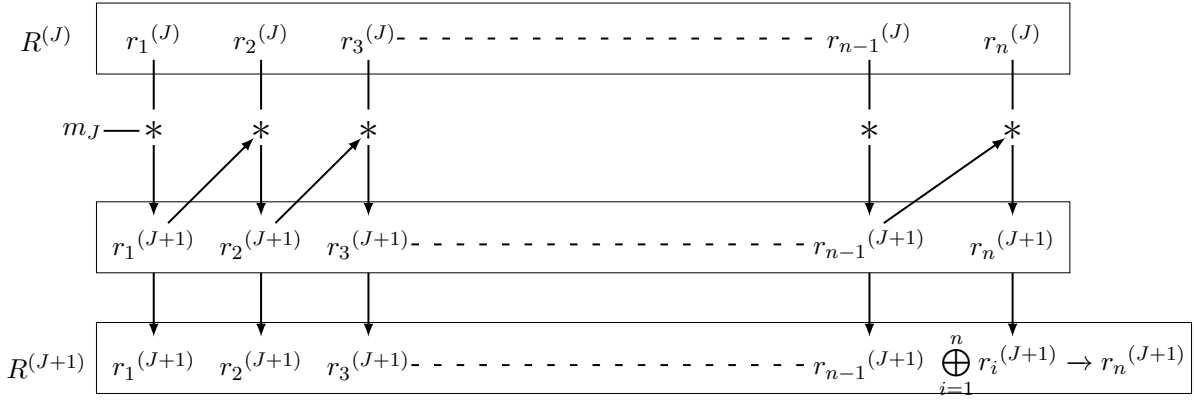

	Now, the encryption scheme proceeds by making use of $\mathfrak{e}$-transformation as defined in Definition \ref{def:e-transformation}. 
	Define the intermediate map $E_{*}:Q\times Q^n\rightarrow Q\times Q^n$
	as 
	\begin{equation}\label{eq:encryptionmap}
		E_{*}(m,(r_1,r_2,\dots,r_n))=\left(r_n*(r_{n-1}*(\dots*(r_1*m)\cdots)),(r_1',r_2',\dots,r_n')\right).
	\end{equation}
	Here, $(r_1',r_2',\dots,r_n')=F_{m}^{n}(r_1,r_2,\dots,r_n)$.  For the pictorial representation, refer Figure \ref{fig:transformvector}.
	
	The encryption process runs sequentially by first applying $E_*$ on $m_1$ using the initial vector. Subsequently, the initial vector gets updated and using this together with $m_2$, the $E_*$ results in second block of ciphertext. Similarly, the initial vector is updated and utilized to encrypt $m_i$ till the complete message is encrypted.
	We say that the ciphertext corresponding to $M$ is given by the vector $\left(E_*(m_1),E_*(m_2),\dots,E_*(m_l)\right)$, where by $E_*(m_i)$ we consider only the first component. 
	This process is explained in Algorithm \ref{algo:encscheme}.
	
\begin{figure}[H]
		\resizebox{\columnwidth}{!}{
			\begin{tikzpicture}
				\draw[white] (0,0) grid (19,7); 
				\node[rectangle,draw, minimum width = 3cm, 
				minimum height = 2cm] (r) at (3.5,4) {$E_*(m_1,R^{(1)})$};
				\node[rectangle,draw, minimum width = 3cm, 
				minimum height = 2cm] (r) at (8,4) {$E_*(m_2,R^{(2)})$};
				\node[rectangle,draw, minimum width = 3cm, 
				minimum height = 2cm] (r) at (12.5,4) {$E_*(m_3,R^{(3)})$};
				\node[rectangle,draw, minimum width = 3cm, 
				minimum height = 2cm] (r) at (18,4) {$E_*(m_l,R^{(l)})$};
				\draw[thick] [-latex] (0.5,4)--(2,4);
				\draw[thick] [-latex] (3.5,6.5)--(3.5,5);
				\draw[thick] [-latex] (8,6.5)--(8,5);
				\draw[thick] [-latex] (12.5,6.5)--(12.5,5);
				\draw[thick] [-latex] (18,6.5)--(18,5);
				\draw[thick] [-latex] (3.5,3)--(3.5,1.5);
				\draw[thick] [-latex] (8,3)--(8,1.5);
				\draw[thick] [-latex] (12.5,3)--(12.5,1.5);
				\draw[thick] [-latex] (18,3)--(18,1.5);
				\draw[thick] [-latex] (4.5,3)--(4.5,1.5)--(5.7,1.5)--(5.7,4)--(6.5,4);
				\draw[thick] [-latex] (9,3)--(9,1.5)--(10.2,1.5)--(10.2,4)--(11,4);
				\draw[thick] [-latex] (13.5,3)--(13.5,1.5)--(14.7,1.5)--(14.7,4)--(15.5,4);
				\draw[thick, dashed] [-latex] (15.5,4)--(16.5,4);
				\node at (0.2,4.1) {$R^{(1)}$};
				\node at (6,4.3) {$R^{(2)}$};
				\node at (10.5,4.3) {$R^{(3)}$};
				\node at (3.5,6.7) {$m_1$};
				\node at (8,6.7) {$m_2$};
				\node at (12.5,6.7) {$m_3$};
				\node at (18,6.7) {$m_l$};
				\node at (3.5,1.3) {$c_1$};
				\node at (8,1.3) {$c_2$};
				\node at (12.5,1.3) {$c_3$};
				\node at (18,1.3) {$c_l$};
			\end{tikzpicture}
		}
		\caption{Encryption algorithm}
		\label{fig:encscheme}
	\end{figure}
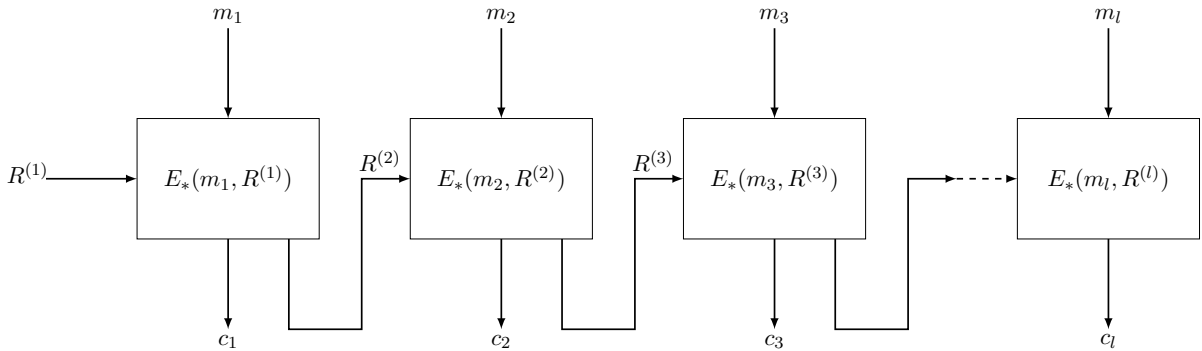

	To encrypt the $i^{th}$ block of message $M$ we need two inputs, i.e. a message block $m_i$ and a random vector $R^{(i)}$, after encrypting the message block $m_i$ we get output $(c_i,R^{(i+1)})$.  For the pictorial representation of encryption scheme refer Figure \ref{fig:encscheme} and Algorithm \ref{algo:encscheme}.
	
	\begin{algorithm}[H]
		\caption{Encryption algorithm}
		\label{algo:encscheme}
		
		\begin{algorithmic}[1]
			\Require {An $n \times n$ sized Latin square $*$ over $Q$, an initial (or, leader) random  vector $R=(r_1,r_2,\dots,r_n)\in Q^n$,  and a message $M=(m_1,m_2,\dots,m_l)\in Q^l=\mathbb{F}_2^{kl}$. } 
			
			\For{$j=1$ to $l$}
			\State $k_{0,j}\leftarrow m_j$ \Comment{Assign $(k_{0,1},k_{0,2},\dots,k_{0,l})=(m_1,m_2,\dots,m_l)$ }
			\EndFor
			
			\For{$i=1$ to $n$}
			\State $k_{i,0}\leftarrow r_i$ \Comment{Assign $(k_{1,0},k_{2,0},\dots,k_{n,0})=(r_1,r_2,\dots,r_n)$ }
			\EndFor
			
			\For{$j=1$ to $l$}
			\For{$i=1$ to $n$}
			\State $k_{i,j}\leftarrow k_{i,j-1}*k_{i-1,j}$ \Comment{Compute $k_{1,j},k_{2,j},\dots, k_{n,j}$}
			\EndFor
			\State $c_j\leftarrow k_{n,j}$  \Comment{ Compute the ciphertext block $c_j$ as $k_{n,j}$}
			\State $k_{n,j}\leftarrow k_{1,j}\oplus k_{2,j}\oplus\dots \oplus k_{n,j}$ \Comment{Update the value of $k_{n,j}$}
			\EndFor	
			\Ensure $C=(c_1,c_2,\dots, c_l)\in Q^l=\mathbb{F}_2^{kl}$
		\end{algorithmic}
		
		
	\end{algorithm}

	Here, the encryption function, i.e. a function from $Q^l\rightarrow Q^l$, can be decomposed into $l$ maps $E_{*}$ each from $Q\times Q^n\rightarrow Q\times Q^n$ and same structure follows for the decryption function as well operation $*$ being replaced with $\backslash$.
	
	\subsection{Decryption Process}\label{dec_sub}
	
	The decryption process takes input a ciphertext $C=(c_1,c_2,\dots,c_l)\in \mathbb{F}_2^{kl}$ along with the initial vector $R^{(1)}=r_1^{(1)},r_2^{(1)},\dots,r_n^{(1)}\in Q^n$. The algorithm requires the multiplication table of the parastrophe $\backslash$ of quasigroup $(Q,*)$ which acts as the secret key. 
	
	For the decryption scheme, we use $\mathfrak{d}$-transformation of quasigroup using Definition  \ref{def:d-transformation}. Define a map 
	$D_{\backslash}:Q\times Q^n\rightarrow Q\times Q^n$ as $D_{\backslash}(c,(s_1,s_2,\dots,s_n))=(r_1\backslash(r_2\backslash(\dots\backslash(r_n\backslash c)\dots)),(s_1',s_2',\dots,s_n'))$. Here, $(s_1',s_2',\dots,s_n')=G_c^n(s_1,s_2,\dots,s_n)$.
	
	Using same definition, 
	transform the vector $S$ after the decryption of each block of ciphertext given the initial vector $S^{(1)}=R^{(1)}=(r_1^{(1)},r_2^{(1)},\dots,r_n^{(1)})$.

	\begin{algorithm}[H]
		\caption{Decryption algorithm}
		\begin{algorithmic}[1]
			\Require An $n\times n$ sized multiplication table of parastrophe $\backslash$ of Latin square $*$ over $Q$, an initial (or, leader) random vector $R=(r_1,r_2,\dots,r_n)\in Q^n$, and a ciphertext $C=(c_1,c_2,\dots,c_l)\in Q^l=\mathbb{F}_2^{kl}$.		
			
			\For{$i=1$ to $n$}
			\State $k_{i,0}\gets r_i$\Comment{Assign $(k_{1,0},k_{2,0},\dots,k_{n,0})=(r_1,r_2,\dots,r_n)$}
			\EndFor
			\For{$j=1$ to $l$}
			\State $k_{n,j}\gets c_j$\Comment{Assign $(k_{n,1},k_{n,2},\dots,k_{n,l})=(c_1,c_2,\dots,c_l)$}
			\EndFor
			\For{$j=1$ to $l$}
			\For{$i=n$ down to 2}
			\State $k_{i-1,j}\leftarrow k_{i,j-1}\backslash k_{i,j}$
			\Comment{Compute $k_{n-1,j},k_{n-2,j},\dots,k_{1,j}$}
			\EndFor
			
			\State $m_j=k_{1,j-1}\backslash k_{1,j}$ \Comment{Compute the message block $m_j=k_{1,j-1}\backslash k_{1,j}$}
			\State $k_{n,j}=k_{1,j}\oplus k_{2,j}\oplus \dots \oplus k_{n,j}$ \Comment{Update the value of $k_{n,j}$}
			\EndFor
			\Ensure $M=(m_1,m_2,\dots, m_l)\in Q^l=\mathbb{F}_2^{kl}$.	
		\end{algorithmic}
		
	\end{algorithm}
	
	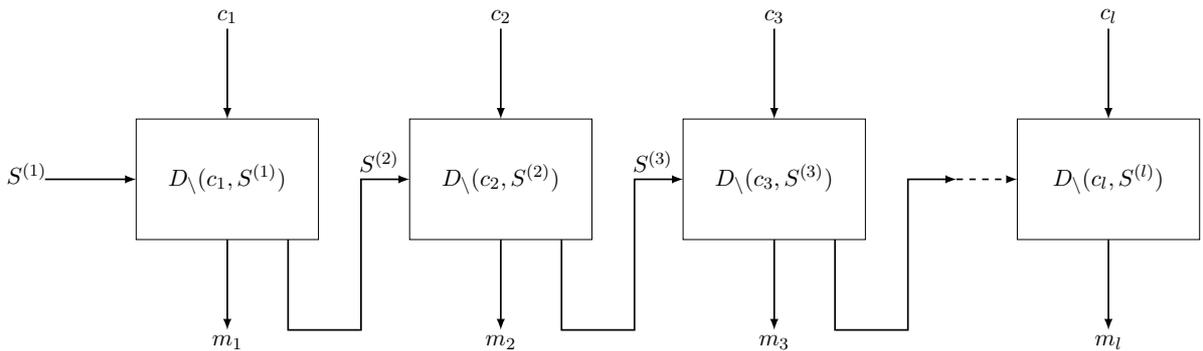
\begin{figure}[H]
		\resizebox{\columnwidth}{!}{
			\begin{tikzpicture}
				\draw[white] (0,0) grid (19,7); 
				\node[rectangle,draw, minimum width = 3cm, 
				minimum height = 2cm] (r) at (3.5,4) {$D_{\backslash}(c_1,S^{(1)})$};
				\node[rectangle,draw, minimum width = 3cm, 
				minimum height = 2cm] (r) at (8,4) {$D_{\backslash}(c_2,S^{(2)})$};
				\node[rectangle,draw, minimum width = 3cm, 
				minimum height = 2cm] (r) at (12.5,4) {$D_{\backslash}(c_3,S^{(3)})$};
				\node[rectangle,draw, minimum width = 3cm, 
				minimum height = 2cm] (r) at (18,4) {$D_{\backslash}(c_l,S^{(l)})$};
				\draw[thick] [-latex] (0.5,4)--(2,4);
				\draw[thick] [-latex] (3.5,6.5)--(3.5,5);
				\draw[thick] [-latex] (8,6.5)--(8,5);
				\draw[thick] [-latex] (12.5,6.5)--(12.5,5);
				\draw[thick] [-latex] (18,6.5)--(18,5);
				\draw[thick] [-latex] (3.5,3)--(3.5,1.5);
				\draw[thick] [-latex] (8,3)--(8,1.5);
				\draw[thick] [-latex] (12.5,3)--(12.5,1.5);
				\draw[thick] [-latex] (18,3)--(18,1.5);
				\draw[thick] [-latex] (4.5,3)--(4.5,1.5)--(5.7,1.5)--(5.7,4)--(6.5,4);
				\draw[thick] [-latex] (9,3)--(9,1.5)--(10.2,1.5)--(10.2,4)--(11,4);
				\draw[thick] [-latex] (13.5,3)--(13.5,1.5)--(14.7,1.5)--(14.7,4)--(15.5,4);
				\draw[thick, dashed] [-latex] (15.5,4)--(16.5,4);
				\node at (0.2,4.1) {$S^{(1)}$};
				\node at (6,4.3) {$S^{(2)}$};
				\node at (10.5,4.3) {$S^{(3)}$};
				\node at (3.5,6.7) {$c_1$};
				\node at (8,6.7) {$c_2$};
				\node at (12.5,6.7) {$c_3$};
				\node at (18,6.7) {$c_l$};
				\node at (3.5,1.3) {$m_1$};
				\node at (8,1.3) {$m_2$};
				\node at (12.5,1.3) {$m_3$};
				\node at (18,1.3) {$m_l$};
			\end{tikzpicture}
		}
		\caption{Decryption algorithm}
		\label{fig:decscheme}
	\end{figure}
	
	Similar to encryption function, here the decryption function, i.e. a function from $Q^l\rightarrow Q^l$ can be decomposed into $l$ maps $D_{\backslash}$ each from $Q\times Q^n\rightarrow Q\times Q^n$ and to decrypt the $i^{th}$ block of ciphertext $C$ we compute $D_{\backslash}(c_i,S^{i})=(m_i,S^{i+1})$.  
	
	\section{Security Analysis of the Proposed Cryptosystem SEBQ }\label{sec:security analysis of MOBQ}

	This section contains the security analysis  of the proposed scheme SEBQ.  For the security analysis, we apply the  conceptual framework of Bellare et al. \cite{12} for IND-CPA and IND-CCA security assessments. 
	We will show  that the SEBQ is IND-CPA secure and subsequently, through the application of an unbalanced Feistel transformation, we establish its IND-CCA2 security.

	\begin{theorem}\label{thm:indcpa}
		The symmetric encryption scheme SEBQ as defined in Section \ref{sec:mobq} based on quasigroup is immune to chosen-plaintext attacks. In other words, any adversary having access to encryption oracle, has negligible advantage in distinguishing correct from his chosen two plaintexts, corresponding to any arbitrary ciphertext.
		\begin{proof}
			In order to prove that the symmetric encryption SEBQ scheme is secure against CPA attack heuristically, first, we consider an experiment (${\bf Exp}$) conducted by adversary $\mathcal{A}$ which generates a key, two plaintexts and a ciphertext of randomly selected one plaintext out of those. The experiment outputs 1 if $\mathcal{A}$ is able to identify the correct plaintext. Algorithmically, it is explained below:
			
			\begin{algorithm}[H]
				\caption{ ${ \bf Exp}_{\mathcal{A},SEBQ}^{cpa}(\text{security parameter})$}
				\begin{algorithmic}[1]
					\State The adversary $\mathcal{A}$ is given input $1^{*}$ and has access to the oracle $\mathcal{E}_{{K}}(\cdot)$. It then produces a pair of messages $m_0$, $m_1$ of the equal length.
					
					\State A random bit $b\in\{0,1\}$ is selected, and subsequently, a ciphertext $c\leftarrow \mathcal{E}_{{K}}(m_b)$ is generated and given to adversary $\mathcal{A}$.
					
					\State The adversary $\mathcal{A}$ continues to have the oracle access $\mathcal{E}_{{K}}(\cdot)$, and following a series of computations, produces an outputs bit $b{'}$.
					
					\State The output of the experiment is  1 if $b{'}=b$, and 0 otherwise. In the former case, we say that the adversary $\mathcal{A}$ succeeds.
				\end{algorithmic}
			\end{algorithm}
			\noindent Suppose adversary chooses plaintexts $m_0$ and $m_1$ from $Q$ and an initial vector $R\in Q$; and gets a ciphertext $C$ of one of the plaintexts. 
			We consider the cases where adversary can query repeated or non-repeated messages to the oracle.
			\begin{itemize}
				\item {\bf Adversary queries the repeated messages with different initial vectors (IVs)}: Since, the encryption scheme SEBQ is probabilistic (not deterministic) and adversary $\mathcal{A}$ queries the repeated message as many times as it wants. 
				Suppose adversary $\mathcal{A}$ repeatedly queries $m_0\in Q$ with initial vectors $R$, with $R'\neq R$ to oracle $\mathcal{E}_{{K}}(\cdot)$. Then, the adversary can compute the complete column of Latin square. Eventually,  ${\bf Exp}_{\mathcal{A},SEBQ}^{cpa}=1$, i.e. Adversary is able to distinguish the plaintext given its ciphertext with probability 1. Thus, the encryption scheme is insecure. However, this kind of liberty is not given to adversary. So we consider the other case only.

				\item {\bf Adversary is not allowed to query the repeated messages}: The main idea for adversary to be successful is to construct the Latin square using the adaptive interaction with the $\mathcal{E}_{{K}}(\cdot)$ oracle. 
				Adversary queries $m_i$, $i\notin \{0,1\}$ to the oracle and get different ciphertexts. This results in detection of whole Latin square except two columns corresponding to $m_0$ and $m_1$ as given in Figure 4. 
				However, even with this knowledge adversary is unable to identify whether the given ciphertext corresponds to which message $m_0$ or $m_1$ as the $R$-th row can have $C$ in either column with equal probability, i.e. $Pr[b=0]=Pr[b=1]=\frac{1}{2}$, which implies ${\bf Adv}_{\mathcal{A}}^{ind-cpa}=0$.

				\begin{figure}[H]
					\centering
					\begin{tikzpicture}[scale=0.4]
						\draw[white] (0,0) grid (14,8); 
						\draw[thick] (2,8)--(2,0.5);
						\draw[thick] (5,6.5)--(5,0.5);
						\draw[thick] (6.5,6.5)--(6.5,0.5);
						\draw[thick] (9.5,6.5)--(9.5,0.5);
						\draw[thick] (11,6.5)--(11,0.5);
						\draw[fill, lightgray!50] (2,6.5)--(5,6.5)--(5,0.5)--(2,0.5)--(2,6.5);
						\draw[fill, lightgray!50] (6.5,6.5)--(9.5,6.5)--(9.5,0.5)--(6.5,0.5)--(6.5,6.5);
						\draw[fill, lightgray!50] (11,6.5)--(14,6.5)--(14,0.5)--(11,0.5)--(11,6.5);
						\draw[thick] (2,6.5)--(2,0.5);
						\draw[thick] (5,6.5)--(5,0.5);
						\draw[thick] (6.5,6.5)--(6.5,0.5);
						\draw[thick] (9.5,6.5)--(9.5,0.5);
						\draw[thick] (11,6.5)--(11,0.5);
						\draw[thick] (0.5,6.5)--(14,6.5);
						\node[scale=3] at (1.5,7.1) {\tiny \scalebox{.8}{*}};
						\node[scale=1] at (1.5,3.5) {\scalebox{.8}{$R$}};
						\node[scale=2] at (5.75,7.1) {\tiny\scalebox{.8}{$m_0$}};
						\node[scale=2] at (10.2,7.1) {\tiny \scalebox{.8}{$m_1$}};
					\end{tikzpicture}
					
					\caption{Construction of Latin square during IND-CPA attack}
					\label{fig:IND-CPAattack}
				\end{figure}
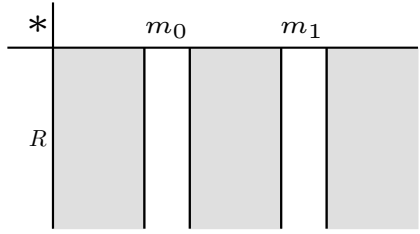
				
			\end{itemize}
		\end{proof}
	\end{theorem}

	\begin{prop}\label{prop:indcca}
		The  SEBQ encryption scheme in not IND-CCA2 secure.
		\begin{proof}
			Suppose adversary $\mathcal{A}$ has access to both encryption $\mathcal{E}_{{K}}(\cdot)$ and decryption $\mathcal{D}_{{K}}(\cdot)$ oracle. 
			Upon querying $\mathcal{D}_{{K}}(\cdot)$ for $C'\neq C$, the complete Latin square can be computed expect the places where $C$ is present. Using the Definition \ref{def:latin square} of Latin square, those left out places can be easily filled. Thus,  ${\bf Adv}_{SEBQ}^{ind-cca}(\mathcal{A})=1$, hence the proposition follows.
		\end{proof}
	\end{prop}
	Similar to the above, it can be readily observed that the same result holds for Left-or-Right indistinguishability. 
	Thus, we aim on applying some transformations to modify the scheme to get indistinguishability under CCA2 attacks.
	
	\subsection{Unbalanced Feistel transformation of SEBQ encryption scheme}
	We transform SEBQ  scheme using unbalanced Feistel transformation \cite{desai} and secure it against adaptive chosen ciphertext attacks. Additionally, using this transformation, the cryptosystems achieves another security goal of non-malleability as introduced by Dolev et al. \cite{dolev}.

	Let SEBQ $=\{\mathcal{K},\mathcal{E},\mathcal{D}\}$ be a symmetric encryption scheme described in Section \ref{sec:mobq}. We proved that it is secure against IND-CPA attack in Theorem \ref{thm:indcpa} and subsequently, in Proposition \ref{prop:indcca} we have shown that it is not IND-CCA2 secure. So, to shield SEBQ scheme against IND-CCA2 attack we transform the SEBQ into $\widetilde{SEBQ}$ using unbalanced Feistel transformation. 
	
	Let $M$ be a variable-length input pseudorandom function, which takes input of any pre-specified length  and outputs some fixed length vector. 
	Similarly, let $G$ be a variable-length output pseudorandom function.
	Our transformation yields a better version of work carried out by  \cite{desai, bellare_07}. 
	
	Let us first recall that the fixed-length pseudorandom functions come from a family of keyed multi-set in which functions have fixed (finite) domain and range corresponding to a key. It must be noted that a pseudorandom function is indistinguishable from a random function within the same domain and range. 
	Variable-length input pseudorandom functions accepts an input of arbitrary and variable lengths, generating output of a fixed length. Most of the MACs are examples for this family. The variable-output pseudorandom functions takes fixed length input and outputs a variable length vector of user's choice. 
	
	The transformation that we apply to our encryption scheme that makes it secure against CCA depends on security of these variable-length output pseudorandom function.

	Let $k_0$ be the positive integer.
	The transformation of $SEBQ$ to $\widetilde{SEBQ}$ is describe by following algorithm:
	
	\begin{algorithm}[H]
		\caption{Transformed symmetric scheme $\widetilde{SEBQ}$}\label{se_trans}
		\begin{algorithmic}[1]
			\State Consider a quasigroup $(Q,*)$ of order $2^k$, assuming $Q=\mathbb{F}_2^k$, an initial vector $R=(r_1,r_2,\dots,r_{k_0})\in Q^{k_0}$, and message $M=(m_1,m_2,\dots,m_l)\in Q^l$.
			
			\State Consider a variable-length output pseudorandom function $G:\mathbb{F}_2^{k_0}\rightarrow(\mathbb{F}_2^{k})^*$.
			
			\State In addition to quasigroup, the key generation algorithm also outputs a positive integer $a>1$ for variable-length output pseudorandom function $G$.
			
			\State The transformed scheme modifies the function $E_*$ as described in Subsection \ref{enc_sub} to $\widetilde{E}_*$ as: 
			\begin{equation}
				\widetilde{E}_*\left(m_i,R^{(i)}\right)={E}_*\left(m_i,G_a(R^{(i)})\right)
			\end{equation}
			
			\State The function $D_\backslash$ as described in Subsection \ref{dec_sub} gets transformed to  $\widetilde{D}_{\backslash}$ as:
			\begin{equation}\label{dec_oracle}
				\widetilde{D}_{\backslash}\left(c_i,S^{(i)}\right)={D}_\backslash\left(c_i,G_a(S^{(i)})\right).
			\end{equation}
		\end{algorithmic}
	\end{algorithm}
	Recall that the encryption scheme $SEBQ$ is insecure to adaptive chosen-ciphertext attacks as it can be easily observed that adversary can distinguish the correct plaintext from out of two, given ciphertext of any one. 
	This distinguisher is possible only when message and initial vector $R$ are both elements of $Q$, but on applying the variable-length output pseudorandom function the length of $G_a(R)$ becomes equal to $a$, which is a positive integer greater than $1$.
	Now upon querying decryption oracle $\widetilde{D}_\backslash(\cdot)$ as described in (\ref{dec_oracle}), it becomes impossible to recover the quasigroup $(Q,*)$. 
	This is due to the increase in length of initial vector that is to be used in generation of Latin square corresponding to $(Q,*)$. As if initial vector was of length $1$, then on querying decryption oracle gives the desired entry of Latin square as explained in Proposition \ref{prop:indcca}. However, this transformation yields in slight increase of number of bit-operations in the encryption scheme, yet it provides security under adaptive chosen-ciphertext attacks.
	With these in-sights, we provide the following theorem.
	\begin{theorem}
		The transformed symmetric encryption scheme $\widetilde{SEBQ}$, based on quasigroup, as described in Algorithm \ref{se_trans} is IND-CCA2 secure.
	\end{theorem}

	\subsection{Randomness testing of SEBQ}
	This section delve into the assessment of the entropy in the symmetric encryption (SEBQ) scheme, as  
	proposed in Algorithm \ref{algo:encscheme}. To accomplish that, we conduct tests using the  NIST test suite \cite{42} and recorded  the outcome into the Tables \ref{table:randomplaintextNISTtestsuite}, \ref{table:randomzerotextNISTtestsuite} and \ref{table:randomonestextNISTtestsuite}. Additionally,  we performed a comparative analysis of the outcomes against  BCWST \cite{38}, INRU \cite{27} in CBC mode and AES-128.
	In this experiment we will examine a 4000-bit long binary sequence  serving as ciphertext. 
	To generate the sequence of ciphertext we utilize a secret key and random initial vector both have 400-bit long sequence. The random initial vector was generated by following the  ANSI x9.31 standard random number generation.  
	These experiments  were carried out   with a  significance level set at $99\%$. Tables \ref{table:randomplaintextNISTtestsuite}, \ref{table:randomzerotextNISTtestsuite} and \ref{table:randomonestextNISTtestsuite} display the success rate  and  consistency of p-values acquired through the execution of the NIST test suite on  three types of plaintexts: 0x00, 0xFF and randomly generated plaintext. We present the  uniformity of  p-values derived from statistical tests by partitioning  the  intervals 0 and 1  into 10 sub-intervals. These p-values are the results of a chi-square test. In our comparative 
	analysis of the  SEBQ encryption scheme  with  BCWST \cite{38}, INRU \cite{27} and AES-128 we analyze the randomness of the  cipher produced by Algorithm \ref{algo:encscheme}. Our analysis revealed that the randomness of SEBQ encryption is comparable with exiting symmetric encryption schemes like AES-128 and in some instances, it surpassed  that of the previous schemes based on quasigroups and AES-128, affirming its superiority   in terms of randomness and security.
	
	In the experiment, we begin by generating three types of plaintext: Random plaintext, one composed solely  of zeroes (0x00) and one composed entirely of ones (0xFF), each with a length of 4000 bits. We then apply  Algorithm \ref{algo:encscheme} to generate ciphertexts corresponding to these  plaintexts. Subsequently, we independently subject  the generated ciphertexts to 
	the NIST test suite and document the resulting  outcomes in their respective tables. 
	
	\begin{table}[H]
		\centering
		\resizebox{\textwidth}{!}
		{\begin{tabular}{|l|ll|ll|ll|ll|}
				\hline
				\multicolumn{1}{|c|}{\multirow{2}{*}{Test}}            & \multicolumn{2}{c|}{SEBQ}        & \multicolumn{2}{c|}{BCWST \cite{38}}                       & \multicolumn{2}{c|}{INRU\cite{27}}                        & \multicolumn{2}{c|}{AES-128}                     \\ \cline{2-9} 
				\multicolumn{1}{|c|}{}                                 & \multicolumn{1}{l|}{Success \%} & p value        & \multicolumn{1}{l|}{Sucess \%}  & p value        & \multicolumn{1}{l|}{Success \%} & p value        & \multicolumn{1}{l|}{Success \%} & p value        \\ \hline
				Frequency                                            & \multicolumn{1}{l|}{100}        & 0.534146         & \multicolumn{1}{l|}{100}        & 0.1025         & \multicolumn{1}{l|}{100}        & 0.1626         & \multicolumn{1}{l|}{99}         & 0.8165         \\ \hline
				Block Frequency                                       & \multicolumn{1}{l|}{99}         & 0.739918         & \multicolumn{1}{l|}{100}        & 0.6371         & \multicolumn{1}{l|}{100}        & 0.4943         & \multicolumn{1}{l|}{99}         & 0.9914         \\ \hline
				Runs                                                   & \multicolumn{1}{l|}{100}        & 0.739918          & \multicolumn{1}{l|}{99}         & 0.1626         & \multicolumn{1}{l|}{100}        & 0.6579         & \multicolumn{1}{l|}{98}         & 0.4190         \\ \hline
				Longest runs of ones in a block                         & \multicolumn{1}{l|}{100}        & 0.534146         & \multicolumn{1}{l|}{100}        & 0.5543         & \multicolumn{1}{l|}{98}         & 0.2896         & \multicolumn{1}{l|}{100}        & 0.9716       \\\hline
				Binary matrix rank                                                   & \multicolumn{1}{l|}{100}        & 0.035174          & \multicolumn{1}{l|}{99}         & 0.0457         & \multicolumn{1}{l|}{99}         & 0.1916         & \multicolumn{1}{l|}{98}         & 0.2896         \\ \hline
				Discrete Fourier transform                             & \multicolumn{1}{l|}{100}        & 0.350485          & \multicolumn{1}{l|}{98}         & 0.4944         & \multicolumn{1}{l|}{98}         & 0.4372         & \multicolumn{1}{l|}{99}         & 0.7197         \\ \hline
				Overlapping template matching                          & \multicolumn{1}{l|}{99}         & 0.739918          & \multicolumn{1}{l|}{100}        & 0.5749         & \multicolumn{1}{l|}{99}         & 0.0711         & \multicolumn{1}{l|}{98}         & 0.6993         \\ \hline
				{NOTM} (149 templates) & \multicolumn{1}{l|}{98--100}    & 0.122325--0.991413  & \multicolumn{1}{l|}{96-100}     & 0.0134--0.9978 & \multicolumn{1}{l|}{96--100}    & 0.0066--0.9963 & \multicolumn{1}{l|}{95--100}    & 0.0034--0.9914 \\ \hline
				Maurer's universal statistical                         & \multicolumn{1}{l|}{100}        & 0.520354         & \multicolumn{1}{l|}{100}        & 0.475          & \multicolumn{1}{l|}{98}         & 0.1025         & \multicolumn{1}{l|}{100}        & 0.8676         \\ \hline
				Linear complexity                                     & \multicolumn{1}{l|}{99}         & 0.350485        & \multicolumn{1}{l|}{99}         & 0.5341         & \multicolumn{1}{l|}{99}         & 0.1025         & \multicolumn{1}{l|}{100}        & 0.9114         \\ \hline
				Serial 1                                               & \multicolumn{1}{l|}{100}        &  0.534146         & \multicolumn{1}{l|}{99}         & 0.5341         & \multicolumn{1}{l|}{99}         & 0.0965         & \multicolumn{1}{l|}{98}         & 0.2133         \\ \hline
				Serial 2                                               & \multicolumn{1}{l|}{99}         & 0.911413         & \multicolumn{1}{l|}{99}         & 0.4011         & \multicolumn{1}{l|}{100}        & 0.5141         & \multicolumn{1}{l|}{99}         & 0.9357         \\ \hline
				Approximate entropy                                  & \multicolumn{1}{l|}{100}        & 0.024011         & \multicolumn{1}{l|}{100}        & 0.0146         & \multicolumn{1}{l|}{100}        & 0.7981         & \multicolumn{1}{l|}{99}         & 0.0220         \\ \hline
				Cumulative sum forward                             & \multicolumn{1}{l|}{100}        & 0.971736         & \multicolumn{1}{l|}{99}         & 0.4373         & \multicolumn{1}{l|}{100}        & 0.0046         & \multicolumn{1}{l|}{99}         & 0.9988         \\ \hline
				Cumulative sum backward                               & \multicolumn{1}{l|}{100}        & 0.965373         & \multicolumn{1}{l|}{100}        & 0.6371         & \multicolumn{1}{l|}{99}         & 0.3345         & \multicolumn{1}{l|}{100}        & 0.6371         \\ \hline
				Random excursions (8 states)                                     & \multicolumn{1}{l|}{98.46--100} & 0.000714--0.524432   & \multicolumn{1}{l|}{98.24--100} & 0.0008--0.5544 & \multicolumn{1}{l|}{98.46--100} & 0.0956--0.9411 & \multicolumn{1}{l|}{98.24--100} & 0.0008--0.5544 \\ \hline
				Random excursions variant (18 states)                              & \multicolumn{1}{l|}{96.34--100} & 0.064321--0.974522 & \multicolumn{1}{l|}{98.24--100} & 0.329--0.9879  & \multicolumn{1}{l|}{95.38--100} & 0.0704--0.9705 & \multicolumn{1}{l|}{98.24--100} & 0.0329--0.9879 \\ \hline
		\end{tabular}}
		\caption{A comparative analysis on the success rates and uniformity of the p-values by running the NIST Test Suite for randomly generated plaintext across SEBQ, BCWST, INRU and AES-128 (where, NOTM--Non-overlapping template matching) }
		
		\label{table:randomplaintextNISTtestsuite}
	\end{table}
	
	\begin{table}[H]
		\resizebox{\textwidth}{!}
		{\begin{tabular}{|l|ll|ll|ll|ll|}
				\hline
				\multicolumn{1}{|c|}{\multirow{2}{*}{Test}} &
				\multicolumn{2}{c|}{SEBQ} &
				\multicolumn{2}{c|}{BCWST \cite{38}} &
				\multicolumn{2}{c|}{INRU \cite{27}} &
				\multicolumn{2}{c|}{AES-128} \\ \cline{2-9} 
				\multicolumn{1}{|c|}{} &
				\multicolumn{1}{l|}{Success \%} &
				p value &
				\multicolumn{1}{l|}{Sucess \%} &
				p value &
				\multicolumn{1}{l|}{Success \%} &
				p value &
				\multicolumn{1}{l|}{Success \%} &
				p value \\ \hline
				Frequency &
				\multicolumn{1}{l|}{99} &
				0.513309 &
				\multicolumn{1}{l|}{99} &
				0.9717 &
				\multicolumn{1}{l|}{100} &
				0.4944 &
				\multicolumn{1}{l|}{97} &
				0.5141 \\ \hline
				Block frequency &
				\multicolumn{1}{l|}{100} &
				0.911413 &
				\multicolumn{1}{l|}{100} &
				0.6579 &
				\multicolumn{1}{l|}{99} &
				0.8513 &
				\multicolumn{1}{l|}{98} &
				0.0135 \\ \hline
				Runs &
				\multicolumn{1}{l|}{98} &
				0.934146 &
				\multicolumn{1}{l|}{98} &
				0.9781 &
				\multicolumn{1}{l|}{99} &
				0.6163 &
				\multicolumn{1}{l|}{97} &
				0.6371 \\ \hline
				Longest run of ones in a block &
				\multicolumn{1}{l|}{100} &
				0.739918 &
				\multicolumn{1}{l|}{100} &
				0.6787 &
				\multicolumn{1}{l|}{99} &
				0.5141 &
				\multicolumn{1}{l|}{99} &
				0.4559 \\ \hline
				Rank &
				\multicolumn{1}{l|}{100} &
				0.122325 &
				\multicolumn{1}{l|}{100} &
				0.9357 &
				\multicolumn{1}{l|}{99} &
				0.5141 &
				\multicolumn{1}{l|}{97} &
				0.0329 \\ \hline
				Discrete Fourier transform  &
				\multicolumn{1}{l|}{100} &
				0.534146 &
				\multicolumn{1}{l|}{100} &
				0.7598 &
				\multicolumn{1}{l|}{99} &
				0.1537 &
				\multicolumn{1}{l|}{95} &
				0.6579 \\ \hline
				Overlapping template matching &
				\multicolumn{1}{l|}{99} &
				0.350485 &
				\multicolumn{1}{l|}{98} &
				0.5749 &
				\multicolumn{1}{l|}{98} &
				0.6163 &
				\multicolumn{1}{l|}{99} &
				0.0167 \\ \hline
				{NOTM} (149 templates)&
				\multicolumn{1}{l|}{96--100} &
				0.017912--0.991468 &
				\multicolumn{1}{l|}{96-100} &
				0.004--0.9943 &
				\multicolumn{1}{l|}{96--100} &
				0.004--0.9978 &
				\multicolumn{1}{l|}{95--100} &
				0.0102--0.9942 \\ \hline
				Maures's universal statistical &
				\multicolumn{1}{l|}{100} &
				0.52032 &
				\multicolumn{1}{l|}{98} &
				0.7981 &
				\multicolumn{1}{l|}{99} &
				0.0668 &
				\multicolumn{1}{l|}{98} &
				0.5141 \\ \hline
				Linear complexity &
				\multicolumn{1}{l|}{99} &
				0.739918 &
				\multicolumn{1}{l|}{99} &
				0.2133 &
				\multicolumn{1}{l|}{99} &
				0.6786 &
				\multicolumn{1}{l|}{99} &
				0.9357 \\ \hline
				Serial 1 &
				\multicolumn{1}{l|}{100} &
				0.350485 &
				\multicolumn{1}{l|}{100} &
				0.1816 &
				\multicolumn{1}{l|}{100} &
				0.3669 &
				\multicolumn{1}{l|}{100} &
				0.2897 \\ \hline
				Serial 2 &
				\multicolumn{1}{l|}{99} &
				0.534146 &
				\multicolumn{1}{l|}{99} &
				0.7981 &
				\multicolumn{1}{l|}{99} &
				0.4559 &
				\multicolumn{1}{l|}{99} &
				0.1626 \\ \hline
				Approximate entropy &
				\multicolumn{1}{l|}{100} &
				0.112300 &
				\multicolumn{1}{l|}{100} &
				0.1154 &
				\multicolumn{1}{l|}{98} &
				0.3041 &
				\multicolumn{1}{l|}{99} &
				0.8514 \\ \hline
				Cumulative sum forward &
				\multicolumn{1}{l|}{98} &
				0.876384 &
				\multicolumn{1}{l|}{98} &
				0.7598 &
				\multicolumn{1}{l|}{98} &
				0.4190 &
				\multicolumn{1}{l|}{99} &
				0.2023 \\ \hline
				Cumulative sum backward &
				\multicolumn{1}{l|}{98} &
				0.986294 &
				\multicolumn{1}{l|}{99} &
				0.5141 &
				\multicolumn{1}{l|}{99} &
				0.9879 &
				\multicolumn{1}{l|}{99} &
				0.7399 \\ \hline
				Random excursions (8 states)&
				\multicolumn{1}{l|}{98.46--100} &
				0.200743--0.764423 &
				\multicolumn{1}{l|}{97.18--100} &
				0.2053--0.8810 &
				\multicolumn{1}{l|}{98--100} &
				0.0965--0.8832 &
				\multicolumn{1}{l|}{97.10--100} &
				0.0151--0.7565 \\ \hline
				Random excursions variant (18 states)&
				\multicolumn{1}{l|}{96.34--100} &
				0.03852--0.974556 &
				\multicolumn{1}{l|}{95.77--100} &
				0.3863--0.9643 &
				\multicolumn{1}{l|}{96--100} &
				0.1223--0.9914 &
				\multicolumn{1}{l|}{94.2--100} &
				0.0151--0.8486 \\ \hline
		\end{tabular}}
		\caption{A comparative analysis on the success rates and uniformity of the p-values by running the NIST Test Suite for plaintext consisting solely of zeros (0x00) across SEBQ, BCWST, INRU and AES-128 (where, NOTM--Non-overlapping template matching)}
		\label{table:randomzerotextNISTtestsuite}
	\end{table}
	
	\subsection{Avalanche criterion}
	We examine the impact of the avalanche effect on the ciphertext produced by Algorithm \ref{algo:encscheme}.  When assessing the security of any symmetric encryption scheme, it is crucial to verify that any modification of input plaintext, random initial vector and key even a single bit leads to a significant alteration in the resulting   ciphertext \cite{41} also known as diffusion in ciphertext.   The concept of  avalanche criterion has been mathematically defined in various ways across different contexts, including block ciphers, Boolean  functions and hash functions.  
	
	Furthermore, we conduct a comparative analysis of the avalanche criterion for SEBQ against the existing schemes like INRU \cite{27}, BCWST \cite{38} and AES-128. Subsequently, present the findings in a tabular format.
	
	\begin{table}[H]
		\centering
		\resizebox{\textwidth}{!}
		{\begin{tabular}{|l|ll|ll|ll|ll|}
				\hline
				\multicolumn{1}{|c|}{\multirow{2}{*}{Test}} &
				\multicolumn{2}{c|}{SEBQ} &
				\multicolumn{2}{c|}{BCWST \cite{38}} &
				\multicolumn{2}{c|}{INRU \cite{27}} &
				\multicolumn{2}{c|}{AES-128} \\ \cline{2-9} 
				\multicolumn{1}{|c|}{} &
				\multicolumn{1}{l|}{Success \%} &
				p value &
				\multicolumn{1}{l|}{Sucess \%} &
				p value &
				\multicolumn{1}{l|}{Success \%} &
				p value &
				\multicolumn{1}{l|}{Success \%} &
				p value \\ \hline
				Frequency     & \multicolumn{1}{l|}{100} & 0.350485 & \multicolumn{1}{l|}{99}  & 0.9558 & \multicolumn{1}{l|}{99}  & 0.6371 & \multicolumn{1}{l|}{100} & 0.1626 \\ \hline
				Block frequency       & \multicolumn{1}{l|}{100} & 0.991468 & \multicolumn{1}{l|}{100} & 0.3345 & \multicolumn{1}{l|}{100} & 0.978  & \multicolumn{1}{l|}{99}  & 0.8677 \\ \hline
				Runs     & \multicolumn{1}{l|}{98}  & 0.350485 & \multicolumn{1}{l|}{100} & 0.4012 & \multicolumn{1}{l|}{99}  & 0.0519 & \multicolumn{1}{l|}{99}  & 0.4373 \\ \hline
				Longest run of ones in a block      & \multicolumn{1}{l|}{100} & 0.911413 & \multicolumn{1}{l|}{100} & 0.3346 & \multicolumn{1}{l|}{99}  & 0.2368 & \multicolumn{1}{l|}{100} & 0.3669 \\ \hline
				Rank     & \multicolumn{1}{l|}{99}  & 0.066882 & \multicolumn{1}{l|}{99}  & 0.1719 & \multicolumn{1}{l|}{99}  & 0.0519 & \multicolumn{1}{l|}{100} & 0.5955 \\ \hline
				Discrete Fourier transform     & \multicolumn{1}{l|}{98}  & 0.035174 & \multicolumn{1}{l|}{97}  & 0.9643 & \multicolumn{1}{l|}{98}  & 0.3505 & \multicolumn{1}{l|}{95}  & 0.6163 \\ \hline
				Maurer's universal statistical      & \multicolumn{1}{l|}{99}  & 0.499323 & \multicolumn{1}{l|}{99}  & 0.5141 & \multicolumn{1}{l|}{99}  & 0.1223 & \multicolumn{1}{l|}{100} & 0.4943 \\ \hline
				Overlapping template matching      & \multicolumn{1}{l|}{99}  & 0.122325 & \multicolumn{1}{l|}{98}  & 0.6163 & \multicolumn{1}{l|}{99}  & 0.9114 & \multicolumn{1}{l|}{99}  & 0.1626 \\ \hline
				NOTM(148 template) &
				\multicolumn{1}{l|}{96--100} &
				0.008879--0.911413 &
				\multicolumn{1}{l|}{96-100} &
				0.0046--0.9915 &
				\multicolumn{1}{l|}{96--100} &
				0.0076--0.9914 &
				\multicolumn{1}{l|}{95--100} &
				0.0179--0.9781 \\ \hline
				Linear complexity       & \multicolumn{1}{l|}{98}  & 0.066882 & \multicolumn{1}{l|}{98}  & 0.6993 & \multicolumn{1}{l|}{100} & 0.1537 & \multicolumn{1}{l|}{100} & 0.9463 \\ \hline
				Serial 1 & \multicolumn{1}{l|}{99}  & 0.534146 & \multicolumn{1}{l|}{98}  & 0.0205 & \multicolumn{1}{l|}{99}  & 0.3505 & \multicolumn{1}{l|}{100} & 0.3505 \\ \hline
				Serial 2 & \multicolumn{1}{l|}{99}  & 0.739918 & \multicolumn{1}{l|}{98}  & 0.1373 & \multicolumn{1}{l|}{99}  & 0.2757 & \multicolumn{1}{l|}{100} & 0.2757 \\ \hline
				Approximate entropy       & \multicolumn{1}{l|}{100} & 0.112300 & \multicolumn{1}{l|}{100} & 0.6993 & \multicolumn{1}{l|}{100} & 0.4749 & \multicolumn{1}{l|}{95}  & 0.3345 \\ \hline
				Cumulative sum forward      & \multicolumn{1}{l|}{98}  & 0.994708 & \multicolumn{1}{l|}{99}  & 0.8832 & \multicolumn{1}{l|}{98}  & 0.4349 & \multicolumn{1}{l|}{100} & 0.7399 \\ \hline
				Cumulative sum backward      & \multicolumn{1}{l|}{98}  & 0.823133 & \multicolumn{1}{l|}{99}  & 0.6579 & \multicolumn{1}{l|}{100} & 0.9558 & \multicolumn{1}{l|}{99}  & 0.6993 \\ \hline
				Random excursions (8 states) &
				\multicolumn{1}{l|}{98.46--100} &
				0.120073--0.876423 &
				\multicolumn{1}{l|}{96.88--100} &
				0.1480--0.8623 &
				\multicolumn{1}{l|}{98.21--100} &
				0.00003--0.9114 &
				\multicolumn{1}{l|}{98.55--100} &
				0.0514--0.8755 \\ \hline
				Random excursions variant (18 states) &
				\multicolumn{1}{l|}{96.34--100} &
				0.08522--0.975564 &
				\multicolumn{1}{l|}{98.44--100} &
				0.0822--0.9114 &
				\multicolumn{1}{l|}{94.64--100} &
				0.0020--0.9558 &
				\multicolumn{1}{l|}{97.1--100} &
				0.0190--0.7887 \\ \hline
		\end{tabular}}
		\caption{A comparative analysis on the success rates and uniformity of the p-values by running the NIST Test Suite  for plaintext consisting solely of ones (0xFF) across SEBQ, BCWST, INRU and AES-128 (where, NOTM--Non-overlapping template matching)}
		\label{table:randomonestextNISTtestsuite}
	\end{table}
	
	\subsubsection{Avalanche analysis by modification in key }
	We examine the variation in ciphertext percentages for SEBQ, as presented in   
	Algorithm \ref{algo:encscheme}, whenever a key alteration occurs, specifically when the Latin square is modified. To assess whether SEBQ fulfills the avalanche criterion, we conducted experiments with 10 randomly generated Latin squares. We encrypted  a randomly generated plaintext, comprising 4000 bits long using each unique secret keys along with the 400-bit long initial vector. The outcomes of the  experiments are mentioned in the Table \ref{table:keyavalancheeffect}.  Comparative analysis   of the SEBQ with INRU \cite{27}, BCWST \cite{38} and AES-128 is mentioned in Table  \ref{table:keyavalancheeffect}.
	
	\begin{table}[H]
		\centering
		
		\begin{tabular}{l l l}
			\hline 
			Encryption scheme & Max $\%$ change in ciphertext & Min $\%$ change in ciphertext\\
			\hline
			BCWST \cite{38} &  50.054 & 49.931 \\
			INRU \cite{27} &	50.164 & 49.822\\
			AES-128  & 	50.046 & 49.937 \\
			SEBQ & {\bf 50.352} & {\bf 49.90}\\	
			\hline
		\end{tabular}
		\caption{Comparative analysis of avalanche effect of secret key}
		\label{table:keyavalancheeffect}
	\end{table}
	
	\subsubsection{Avalanche analysis in ciphertext by changing the random initial vector}
	In this section, we conduct an examination of the impact on the ciphertext of SEBQ whenever a single bit in the initial vector is inverted. Our approach involves employing a  randomly generated $16\times 16$ size Latin square  as the secret key to encrypt a randomly generated plaintext vector consisting of 4000-bits. To assess the avalanche effect of initial vector on the ciphertext, we generate 100 random initial vectors and encrypt the generated plaintext using  each of them.   We scrutinized the avalanche effect at each change in bit positions within the initial vector and show some of the results in the Table \ref{table:avalancheeffectof IV}. The final rows of Table \ref{table:avalancheeffectof IV} furnish  the maximum and minimum avalanche effects of the initial vector on ciphertext. This experimentation is repeated 10 times and the outcomes of avalanche effect being stored in Table \ref{table:avalancheeffectof IV}.
	
	\begin{table}[H]
		\caption{Observation of avalanche effect due to  change in initial vector bit positions }
		
		\scalebox{0.90}{\begin{tabular}{clllllllllll}
				\hline
				\multicolumn{1}{|c|}{\multirow{2}{*}{\begin{tabular}[c]{@{}c@{}}Pos.\end{tabular}}} &
				\multicolumn{11}{c|}{Percentage change in ciphertext with corresponding experiment} \\ \cline{2-12} 
				\multicolumn{1}{|c|}{} &
				\multicolumn{1}{l|}{$1^{st}$} &
				\multicolumn{1}{l|}{$2^{nd}$} &
				\multicolumn{1}{l|}{$3^{rd}$} &
				\multicolumn{1}{l|}{$4^{th}$} &
				\multicolumn{1}{l|}{$5^{th}$} &
				\multicolumn{1}{l|}{$6^{th}$} &
				\multicolumn{1}{l|}{$7^{th}$} &
				\multicolumn{1}{l|}{$8^{th}$} &
				\multicolumn{1}{l|}{$9^{th}$} &
				\multicolumn{1}{l|}{$10^{th}$} &
				\multicolumn{1}{l|}{Avg.} \\ \hline
				1 & 50.600 & 50.175 & 48.800 & 50.775 & 50.249 & 49.350 & 49.650 & 48.500 & 50.349 & 50.200 & 49.865\\
				2 & 49.475 & 51.500 & 49.225 & 50.700 & 51.249 & 48.699 & 48.650 & 49.850 & 50.100 & 50.649 & 50.010\\
				3 & 50.275 & 50.050 & 49.850 & 49.700 & 50.100 & 49.550 & 50.949 & 49.350 & 50.824 & 50.550 & 50.120\\
				4 & 50.349 & 50.625 & 50.800 & 49.875 & 51.500 & 49.650 & 51.375 & 49.100 & 50.700 & 49.575 & 50.355\\
				5 & 50.800 & 50.449 & 49.825 & 49.850 & 49.325 & 50.475 & 50.125 &  50.900 & 50.975 & 49.875 & 50.260\\
				6 & 50.724  & 50.075& 49.500 & 50.149 & 50.749 & 48.675 & 50.025 &  50.375 & 50.449 & 50.550 & 50.127\\
				7 & 50.400 & 49.200 & 49.875 & 49.675 & {\bf 48.350} & 50.525 & 51.425 &  49.350 & 50.400 & 50.575 &49.977\\
				8 & 49.400 & 50.900 & 50.824 & 49.850 & 49.925 & 50.025 & 49.875 & 48.750 & 50.525 & 49.800 & 49.987 \\
				9 & 50.875 & 50.600 & 49.050 & 50.400 & 50.824 & 48.375 & 50.175 &  49.425 & 50.375 & 51.775 & 50.187\\
				10 & 49.500 & 50.775 & 50.375 & 49.450 & 50.075 & 49.800 & 49.875  & 50.224 & 51.200 & 50.224 & 50.150\\
				128 & 50.600 & 50.525 & 50.600 & 50.849 & 50.324 & 51.475 & 50.600   & 49.825 & 50.175 & 49.550 & 50.452 \\
				256 & 49.900 & 50.449 & 51.175 & 50.625 & 50.449 & 48.900 & {\bf 51.824}& 49.900 & 49.250 & 49.875 & 50.235\\
				
				\hline
				\multicolumn{12}{|c|}{\bf Maximum $\%$ change in ciphertext is  51.824}                                           \\ \hline
				\multicolumn{12}{|c|}{\bf Minimum $\%$ change in ciphertext is 48.350}                                           \\ \hline
		\end{tabular}}
		\label{table:avalancheeffectof IV}
		
	\end{table}
	\subsubsection{Avalanche analysis in ciphertext by change in plaintext }
	To examine the influence of modifying a bit in the plaintext on the resultant  ciphertext of SEBQ, we carry out  a sequence of experiments. Specifically, we generate  100 random plaintext, each consisting of a  4000-bit long vector. These plaintexts were then individually encrypted using a randomly generated secret  Latin square of order $16\times 16$ in conjunction with a 400-bit long initial vector. The resulting data is also compared with the existing schemes such as 
	INRU \cite{27}, BCWST \cite{38} and AES-128. The comparative outcomes are presented in  Table \ref{table:Avalanche effect of plaintext}. 
	To ensure statistical significance, we repeated these experiments 
	10 times and record the outcomes of avalanche effect in Table \ref{table:Avalanche effect of plaintext}.
	
	\begin{table}[H]
		\centering
		\caption{Comparative analysis for maximum and minimum  avalanche effect of plaintext}
		
		\begin{tabular}{lll}
			\hline
			Encryption scheme & Max $\%$ change in ciphertext & Min $\%$ change in ciphertext\\
			\hline
			BCWST \cite{38} & {50.041} & {49.959} \\
			INRU \cite{27} & {50.042} & {49.943} \\
			AES-128 & 50.032 & 49.957\\
			SEBQ & {\bf 52.550} & {\bf 48.000}\\
			\hline
		\end{tabular}
		\label{table:Avalanche effect of plaintext}
	\end{table}
	
	\begin{table}[H]
		\caption{Observation of avalanche effect due to  change in plaintext bit positions }
		
		\scalebox{.90}{\begin{tabular}{clllllllllll}
				\hline
				\multicolumn{1}{|c|}{\multirow{2}{*}{\begin{tabular}[c]{@{}c@{}}Pos.\end{tabular}}} &
				\multicolumn{11}{c|}{Percentage change in ciphertext with corresponding experiment} \\ \cline{2-12} 
				\multicolumn{1}{|c|}{} &
				\multicolumn{1}{l|}{$1^{st}$} &
				\multicolumn{1}{l|}{$2^{nd}$} &
				\multicolumn{1}{l|}{$3^{rd}$} &
				\multicolumn{1}{l|}{$4^{th}$} &
				\multicolumn{1}{l|}{$5^{th}$} &
				\multicolumn{1}{l|}{$6^{th}$} &
				\multicolumn{1}{l|}{$7^{th}$} &
				\multicolumn{1}{l|}{$8^{th}$} &
				\multicolumn{1}{l|}{$9^{th}$} &
				\multicolumn{1}{l|}{$10^{th}$} &
				\multicolumn{1}{l|}{Avg.} \\ \hline
				1  & 49.875 & 51.200 & 52.050 & 50.050 & 49.650 & 48.525 & 49.925 & 49.300 & 50.625 & 50.800 & 50.200 \\
				2  & 49.875 & 50.375 & 49.075 & 50.949 & 51.075 & 50.625 & 48.949 & 49.450 & 50.349 & 50.324 & 50.105 \\
				3  & 49.375 & 49.750 & 50.050 & 50.675 & 49.825 & 50.800 & 51.025 & 50.849 & 49.475 & 48.850 & 50.067 \\
				4  & 49.400 & 51.675 & 49.675 & 50.625 & 49.000 & 51.050 & 49.225 & 50.050 & 49.425 & 50.125 & 50.025 \\
				5  & 49.150 & 50.324 & 49.550 & 50.324 & 50.575 & 50.550 & 50.349 & 49.675 & 50.275 & 48.699 & 49.947 \\
				6  & 49.600 & 49.825 & 50.400 & {\bf 48.000} & 50.125 & 50.949 & 51.100 & 51.200 & 49.375 & 51.050 & 50.162 \\
				7  & 49.675 & 49.950 & 49.900 & 49.075 & 49.675 & {\bf 52.550} & 49.950 & 49.700 & 49.625 & 49.375 & 49.847 \\
				8  & 50.324 & 49.525 & 51.200 & 49.025 & 49.700 & 50.149 & 50.075 & 49.775 & 49.950 & 49.920 & 49.965 \\
				9  & 50.025 & 50.000 & 49.925 & 50.224 & 49.525 & 50.975 & 47.500 & 50.300 & 51.200 & 50.525 & 50.020 \\
				10 & 49.425 & 48.725 & 50.775 & 49.250 & 49.500 & 51.100 & 50.324 & 49.775 & 50.600 & 50.675 & 50.015\\
				\hline
		\end{tabular}}
		\label{table:avalancgeeffectofplaintext}
	\end{table}
	
	\section{Analysis of SEBQ scheme}\label{sec:analysisof scheme}
	
	This section is committed  to find out the total number of operations needed for  encryption and decryption of plaintext in the proposed SEBQ scheme as detailed in Section \ref{sec:mobq}. Additionally, we will show that the number of operations require to encrypt and decrypt the plaintext with  specific parameters is more efficient compared to  the existing schemes such as  INRU \cite{27} and BCWST \cite{38}.   
	
	Consider $Q=\mathbb{F}_2^k$ and a Latin square of order $2^k$. Now, let's assume that message $M=m_1,m_2,\dots,m_l\in (\mathbb{F}_2^{k})^l$ is being encrypted with the given Latin square in conjunction with the initial vector $IV=r_1,r_2,\dots,r_n$.
	
	\begin{prop}\label{prop:number of operation for enc scheme}
		The number of operations required to encrypt the given message $M$ using the given Latin square and the initial vector $IV$ is $n+(l-1)(n+k)$.
		\begin{proof}
			The number of operations needed  to encrypt the $i^{th}$ block of message $M$ is $(n+k)$ and for the entire message $M$ is  $(l-1)(n+k)$  operations. Additionally, updating the vector requires performing $n$ xor operations. Therefore, the total number of operations required for the encryption scheme is equal to $n+(l-1)(n+k)$.
		\end{proof}
	\end{prop} 
	
	\begin{prop}\label{prop:operationfor decryption}
		The number of operations required to decrypt the given ciphertext $C$ using the given Latin square and the initial vector $IV$ is $n+(l-1)(n+k)$.
		\begin{proof}
			Proof is similar as above Proposition  \ref{prop:number of operation for enc scheme}.
		\end{proof}
	\end{prop}
	Hence, the total  operations required for the symmetric encryption scheme SEBQ are equal to $2(n+(l-1)(n+k))$ which is proportional to $\mathcal{O}(nl)$.

	Consider a scenarios where a user intends to secure a message $M\in \mathbb{F}_2^{64}$ through encryption and by  using 128-bit long secret key along with 16-bit long initial vector. The number of operations required to encrypt the message $M$ is just 70. In contrast, for the same message and parameter set, the INRU: A quasigroup based lightweight block cipher \cite{27} necessitates 154 operations and to encrypt the same message with same parameter  120 number of operations are needed to perform in the encryption scheme BCWST \cite{38}.  
	In a nutshell, the symmetric encryption scheme SEBQ described in Algorithm \ref{algo:encscheme} demonstrated superior efficiency compared to the lightweight block ciphers proposed in \cite{27}, \cite{38}.

	Suppose $Q=\mathbb{F}_2^4$ and a Latin square of order $m$ has been utilized to encrypt the message $M=m_1,m_2,\dots,m_l\in \mathbb{F}_2^{4l}$ where $m_i\in \mathbb{F}_2^4$ in conjunction with random vector $R=r_1,r_2,\dots,r_8\in \mathbb{F}_2^{32}$  where each $r_i\in \mathbb{F}_2^4$. Suppose adversary $\mathcal{A}$ possesses information about the $128$-bit ciphertext. The number of operations needed to make an educated guess about the plaintext corresponding to the provided ciphertext are detailed  in Proposition \ref{prop:securitylevel} and \ref{prop:256security}. The minimum order of Latin square to provide security against 128-bit and 256-bit known ciphertext attack is given by following Proposition \ref{prop:securitylevel} and \ref{prop:256security}.  
	
	\begin{prop}\label{prop:securitylevel}
		In order to attain 128-bit security against known ciphertext attack in SEBQ, it is imperative that the order of Latin square must be greater than 11.
		\begin{proof}
			According to the Proposition \ref{prop:operationfor decryption} the decryption of a 128-bit ciphertext necessitates 380 operations. 
			In order for an adversary $\mathcal{A}$ to successfully guess the 
			correct plaintext corresponding to given ciphertext it  needed to  perform operations greater than $L(m)\times 380$ where $L(m)$ represents the count of Latin squares of order $m$. To achieve  128-bits security level, parameter  $m$ ought to be selected in a manner that  $L(m)\times 380\ge 2^{128}$ which implies that the parameter $m$ should be greater than 11.  Conclusively, achieving  a 128-bit security level implies that order of Latin square should be greater than 11. 
		\end{proof}  
	\end{prop}
	
	\begin{prop}\label{prop:256security}
		In order to attain 256-bit security against known ciphertext attack  in SEBQ, it is imperative that the order of Latin square should be greater than 13.
		\begin{proof}
			Proof is same as Proposition \ref{prop:securitylevel}.
		\end{proof}
	\end{prop}
	
	\section{Conclusion}\label{sec:conclusion}
	This paper presents  a symmetric encryption scheme (SEBQ) based on the structure of quasigroups and the string transformations. The SEBQ scheme employs a chaining-like mode of operation. In this mode, unlike the CBC, a transformed initial vector is utilized to encrypt each subsequent block of message. We proved that the SEBQ scheme is IND-CPA  secure and further enhanced its security to withstand IND-CCA2 attack through the application of  unbalanced Feistel transformation. To assess the performance and randomness of the algorithm, we subjected to the NIST test suite for statistical analysis and conducted comparative analysis with the existing schemes like INRU \cite{27}, BCWST \cite{38} and AES-128. Additionally, we conducted an avalanche analysis of secret key, initial vector and plaintext to gauge the diffusion properties of the SEBQ encryption scheme, comparing it to INRU \cite{27}, BCWST \cite{38} and AES-128. Furthermore, our research included an analysis of SEBQ computational aspects, determining  the number of operations needed to perform encryption and decryption process. Lastly, we identified  the minimum order of Latin squares necessary to attain 128-bit security level against known-ciphertext attack.

	The $\mathfrak{e}$ and $\mathfrak{d}$-transformations as defined in Definition \ref{def:e-transformation} and \ref{def:d-transformation} can play a crucial role in improving the security of  cryptographic algorithms based on quasigroups like SEBQ. This augmentation occurs by employing   more than one quasigroup and more than one round with different initial vectors  \cite{38}. Furthermore,  the proposed design's security  can be evaluated by experimenting with various quasigroups to tailor the algorithms for specific applications.


\end{document}